\documentclass{article} 
\usepackage{colm2024_conference}

\usepackage{amsmath,amsfonts,bm}

\def\eqref#1{equation~\ref{#1}}

\def\1{\bm{1}}

\DeclareMathAlphabet{\mathsfit}{\encodingdefault}{\sfdefault}{m}{sl}
\SetMathAlphabet{\mathsfit}{bold}{\encodingdefault}{\sfdefault}{bx}{n}

\usepackage[T1]{fontenc}
\usepackage{hyperref}
\usepackage{url}
\usepackage{booktabs} 
\usepackage{multirow}
\usepackage{enumitem}
\usepackage{amsmath} 
\usepackage{natbib}
\hypersetup{colorlinks=true, allcolors=blue}
\usepackage{caption}
\usepackage{pifont}
\usepackage{listings}
\usepackage{multirow}
\usepackage[table]{xcolor}
\usepackage{arydshln}
\usepackage{xcolor}
\usepackage{titletoc}
\usepackage{subcaption}
\usepackage{wrapfig}
\usepackage{algorithm}
\usepackage{algorithmic}
\usepackage{amssymb}

\floatname{algorithm}{Algorithm}

\title{PCTD: Preference-Guided Counterfactual Task Decomposition for Agent Tool Retrieval}

\author{
Chu Zhao$^{1}$\thanks{Equal contribution.},
Lei Tang$^{2*}$,
Minghang Li$^{3}$,
Jianzhe Zhao$^{1}$,
Guibing Guo$^{1}$\thanks{Corresponding authors.},
Zhengzong Chen$^{2\dagger}$,
Yuanyuan Zhao$^{2}$,
Fei Huang$^{2}$\\
$^{1}$Northeastern University\\
$^{2}$Honor Device Co., Ltd\\
$^{3}$Beijing University of Posts and Telecommunications
}

\begin{document}
\maketitle
\vspace{-30pt}
\begin{center}
    \ghlink
\end{center}

\begin{abstract}
Task decomposition aims to transform ambiguous instructions into executable atomic subtasks, thereby guiding high-precision tool retrieval. However, our analysis reveals that directly adopting tool retrieval metrics, i.e., Recall or NDCG, as rewards for task decomposition can easily induce reward hacking in reinforcement learning-based methods. Specifically, models tend to maximize retrieval matching through strategies such as repetitive decomposition. This spurious correlation between the shallow features of decomposition results and retrieval metric impairs generalization in Out-of-Domain (OOD) scenarios involving unseen tools.
To address this issue, we propose PCTD, a \textbf{P}reference-guided \textbf{C}ounterfactual \textbf{T}ask \textbf{D}ecomposition framework. PCTD quantifies the marginal causal gain of decomposition on retrieval ranking through a counterfactual reward, thereby cutting off spurious correlations at their source. Meanwhile, it introduces a preference reward to impose fine-grained structural supervision on logical coherence and atomicity, encouraging the model to generate high-quality decompositions. In addition, we construct MTDTool, the task decomposition benchmark specifically designed for mobile multi-turn interactions. Extensive experiments demonstrate that PCTD alleviates repetitive decomposition and surpasses SOTA methods in retrieval, decomposition quality, and OOD generalization.
\begin{figure*}[hb]
\centering

\begin{subfigure}{0.32\textwidth}
\centering
\includegraphics[width=\linewidth]{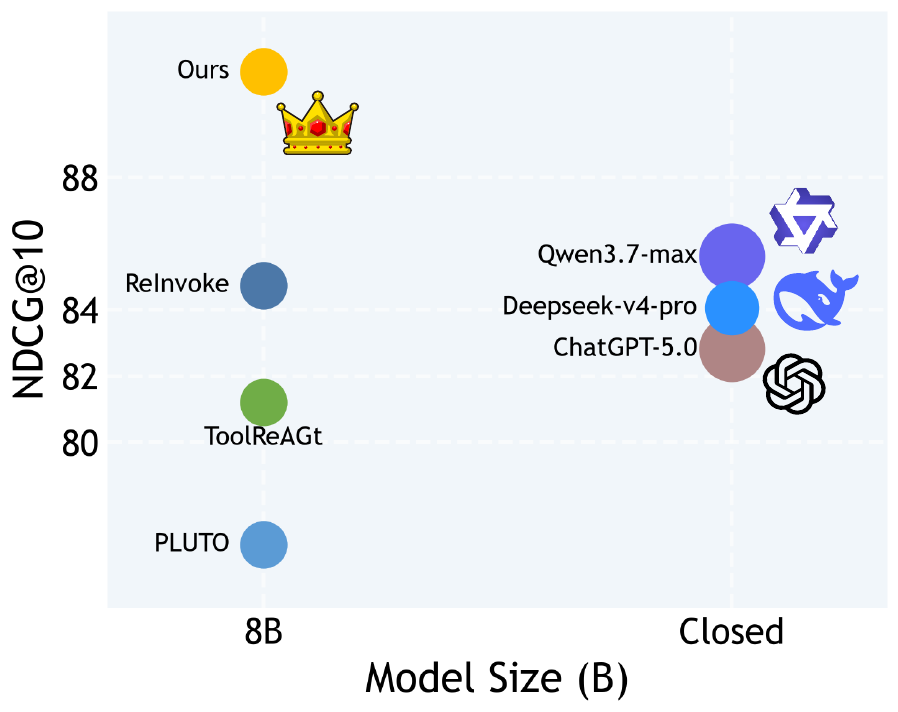}
\end{subfigure}
\hfill
\begin{subfigure}{0.31\textwidth}
\centering
\includegraphics[width=\linewidth]{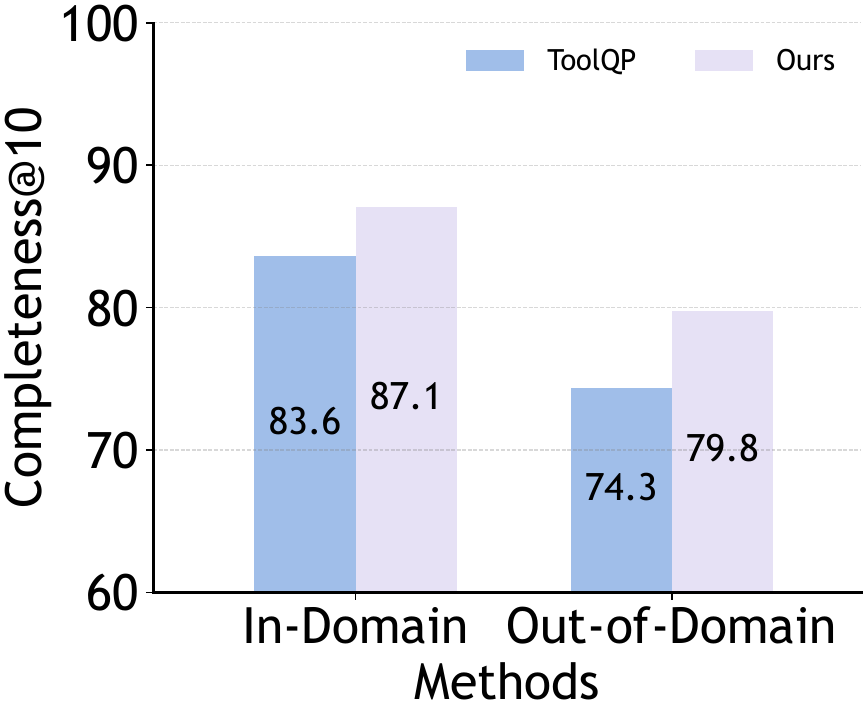}
\end{subfigure}
\hfill
\begin{subfigure}{0.34\textwidth}
\centering
\includegraphics[width=\linewidth]{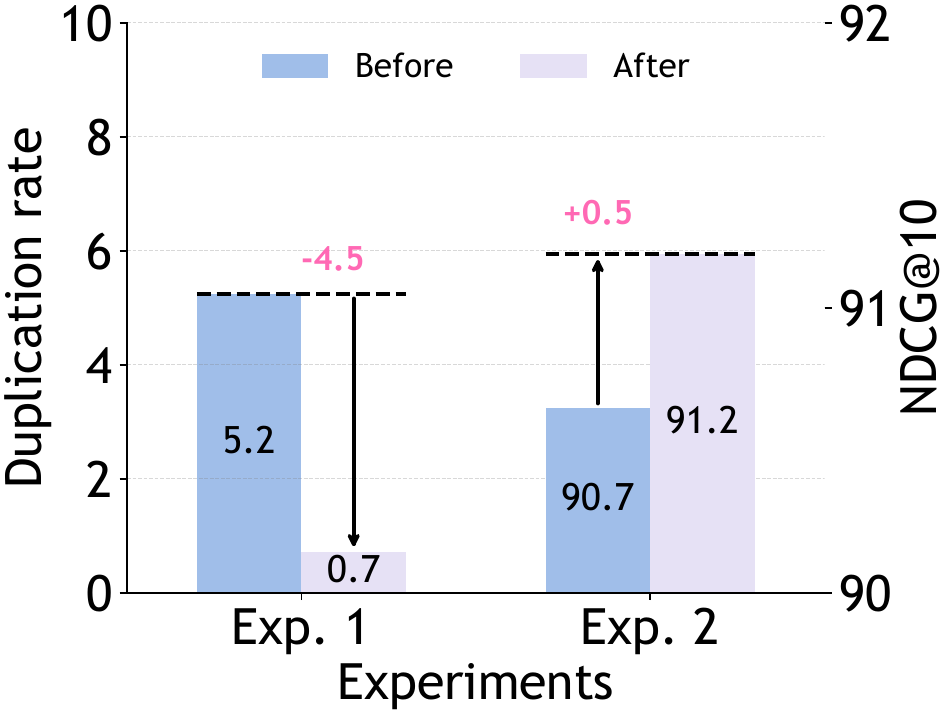}
\end{subfigure}
\caption{Overall performance and empirical study results. Middle: Performance of different models on In-Domain and OOD settings on MTDTool. Right: Model performance as a function of the decomposition repetition rate.
Exp. 1 denotes the repetition rate of task decomposition results before and after applying the preference model, while Exp. 2 represents the change in evaluation metrics before and after applying the preference model.}
\label{fig: motivation}
\end{figure*}
\end{abstract}


\section{Introduction}

Large language model~\citep{zhao2026survey,naveed2025comprehensive, wang2026survey} driven agents are evolving into intelligent entities capable of autonomously decomposing tasks and invoking tools to achieve complex goals~\citep{huang2024understanding,xu2024enhancing}. In multi-turn mobile interaction scenarios, user requests typically rely on dialogue history, coreference relationships, and implicit slots, where a single input is often insufficient to fully describe the objective. This necessitates that agents transform ambiguous instructions into semantically clear, well-bounded, and executable atomic sub-tasks to guide high-precision tool retrieval. Early prompt engineering based methods, such as ReAct~\citep{yao2022react}, ToolReAGt~\citep{brauns2025toolreagt}, Q2E~\citep{jagerman2023query}, and PLUTO~\citep{huang2024planning}, achieved progress in few-shot scenarios through explicit chains of thought or structured plans. However, due to the lack of end-to-end optimization mechanisms and heavy reliance on intrinsic knowledge, these methods are prone to decomposition deviations caused by cumulative errors when facing unseen tools or long-range dependencies. To overcome this limitation, reinforcement learning (RL) based end-to-end methods like ToolQP~\citep{fang2026beyond} incorporated downstream tool retrieval metrics (e.g., Recall@K, NDCG@K) into mixed reward signals. Although these approaches enhance final retrieval performance, they fundamentally rely on static outcome-oriented supervision and suffer from two core limitations.

First, directly employing retrieval metrics as rewards is highly susceptible to \textbf{reward hacking}. Models tend to maximize matching scores through superficial strategies such as repetitive decomposition, leading to a spurious correlation between the shallow lexical features of decomposition results and retrieval outcomes. This severely impairs generalization in Out-of-Domain (OOD) scenarios involving unseen tools. Second, there is a lack of preference modeling for decomposition structure. plan-level similarity-based rewards (e.g., the reward signals utilized in ToolQP) fail to constrain structural quality dimensions such as logical coherence, context faithfulness, and atomicity, further exacerbating the issue of repetitive decomposition. Empirical results in Figure~\ref{fig: motivation} validate this phenomenon: existing methods perform reasonably well on in-distribution data but suffer significant performance degradation on OOD datasets involving unseen tools (e.g., ToolQP drops from 83.6 to 74.3). Meanwhile, a clear negative correlation is observed between repetition rate and performance (as the repetition rate decreases from 5.2\% to 0.7\%, performance improves to 91.2). Therefore, mitigating the spurious correlations and repetitive decomposition induced by reward hacking is critical for improving model performance.

To address these dual challenges, we propose PCTD, a joint optimization framework for task decomposition and tool retrieval tailored for mobile multi-turn interactions. This framework introduces a preference reward $R_{\text{pref}}$, which leverages a trained reward model to score candidate decompositions against human references conditioned on context, providing fine-grained structural quality supervision via relative score differences; simultaneously, it incorporates a counterfactual reward $R_{\mathrm{cf}}$ that uses the retrieval results of the raw query as a baseline to quantify the marginal gain of the decomposition strategy on tool ranking, thereby offering causal feedback for decomposition behavior.
Furthermore, existing benchmarks such as HammerBench~\citep{wang2025hammerbench} and ToolBench~\citep{qin2024toolllm} lack fine-grained process-level annotations, and there is currently no dataset specifically designed for evaluating the task decomposition process, we constructed MTDTool. This benchmark is designed for complex multi-turn interaction scenarios in mobile environments and provides fine-grained evaluation support. Unlike existing datasets that primarily focus on single-turn instructions or final execution outcomes, MTDTool automatically annotates dialogue state evolution, context-aware intent rewriting, and atomic task sequences for each conversational turn. The dataset highlights the dynamic evolution of intents across multi-turn dialogues and cross-application collaboration, providing a standardized basis for quantifying the impact of decomposition quality on retrieval effectiveness.

The contributions of this paper are summarized as follows:
\ding{182} \textit{\textbf{Insight.}} We reveal the dual limitations of existing RL-based agents in task decomposition: (i) directly using retrieval metrics as rewards is prone to reward hacking, severely impairing OOD generalization with unseen tools; and (ii) the lack of structural preference modeling further exacerbates repetitive decomposition.
\ding{183} \textit{\textbf{Method.}} We propose PCTD, a joint optimization framework integrating counterfactual reasoning with preference modeling. It leverages counterfactual rewards to quantify the marginal causal gain of decomposition on retrieval ranking to cut off spurious correlations, while introducing preference rewards to enforce fine-grained structural supervision on logical coherence and atomicity.
\ding{184} \textit{\textbf{Dataset.}} We release MTDTool, the fine-grained benchmark for task decomposition in mobile multi-turn interactions. Generated via a state-machine-driven pipeline, it provides multi-level annotations for dialogue state evolution, intent rewriting, and atomic task sequences, filling the gap in process-oriented evaluation.
\ding{185} \textit{\textbf{Evaluation.}} Extensive experiments on MTDTool and mainstream benchmarks demonstrate that PCTD significantly outperforms SOTA methods in tool retrieval, decomposition quality, and OOD generalization.

\section{Preliminary}
\subsection{Problem Formulation: Multi-Turn Task Decomposition and Tool Retrieval}
We formulate multi-turn agent interaction as a sequential decision process. At turn $t$, the agent receives a user query $q_t$ and the dialogue history $H_t=\{q_1,o_1,\ldots,q_{t-1},o_{t-1}\}$, forming the contextual input $x_t=(H_t,q_t)$. Given a large-scale tool library $\mathcal{T}=\{a_1,\ldots,a_M\}$ containing $M$ APIs, the agent employs a policy model $\pi_\theta$ to generate a natural-language task decomposition sequence $y_t\sim\pi_\theta(\cdot\mid x_t)$. The generated sequence is parsed into a set of semantically clear atomic subtasks $\mathcal{A}_t=\{\tau_1,\ldots,\tau_L\}$, which aims to mitigate semantic ambiguity, implicit slots, and cross-turn coreference in direct retrieval.
Given a top-$K$ tool retriever $R_K(\cdot)$, each atomic subtask is used to retrieve candidate tools, and the retrieved lists are merged, deduplicated, and truncated into a decomposition-aware ranked list:
\begin{equation}
\mathcal{L}^K_t
=
\operatorname{TopK}
\left(
\operatorname{Dedup}
\left(
\operatorname{Merge}
\left(
R_K(\tau_1),\ldots,R_K(\tau_L)
\right)
\right)
\right).
\end{equation}
Let $\mathcal{G}_t\subseteq\mathcal{T}$ denote the ground-truth tool set required for the contextual input $x_t$. Our objective is to optimize the policy $\pi_\theta$ such that the generated decomposition $y_t$ maximizes the ranking consistency between the retrieved list $\mathcal{L}^K_t$ and the ground-truth tool set $\mathcal{G}_t$.

\subsection{Group Relative Policy Optimization}
To achieve efficient policy optimization without training an independent value network (Critic), we adopt Group Relative Policy Optimization (GRPO)~\citep{guo2025DeepSeek}. For each input $x_t$, GRPO samples a group of $B$ candidate outputs $\{y_1, \dots, y_B\}$ from the old policy $\pi_{\theta_{old}}$ and computes their corresponding reward scores $\mathbf{R} =\{R_1, \dots, R_B\}$. The advantage function for each sample is estimated via group-wise normalization: $A_i = \frac{R_i - \text{mean}(\mathbf{R})}{\text{std}(\mathbf{R}) + \epsilon}$,
 This approach leverages relative performance within the group to replace absolute value estimation, significantly reducing memory overhead and enhancing training stability. The optimization objective is defined as:
\begin{equation}
\begin{aligned}
    J_{GRPO}(\theta) = \mathbb{E}_{x \sim \mathcal{D}, \{y_i\} \sim \pi_{\theta_{old}}} \\\Bigg[ \frac{1}{B} \sum_{i=1}^B \bigg( &\min\left(r_i(\theta)A_i, \text{clip}(r_i(\theta), 1-\epsilon, 1+\epsilon)A_i\right) 
    - \beta D_{KL}(\pi_\theta || \pi_{ref}) \bigg) \Bigg],
\end{aligned}
\label{eq: grpo}
\end{equation}
where $r_i(\theta) = \frac{\pi_\theta(y_i \mid x_t)}{\pi_{\theta_{old}}(y_i \mid x_t)}$ is the importance sampling ratio, and $D_{KL}$ denotes the KL divergence penalty relative to a reference model $\pi_{ref}$. In this work, we integrate the proposed Counterfactual NDCG Reward into this framework to drive the targeted evolution of task decomposition strategies.

\section{Method}
\label{methods}

\begin{figure}[htbp]
    \centering
\includegraphics[width=1\textwidth]{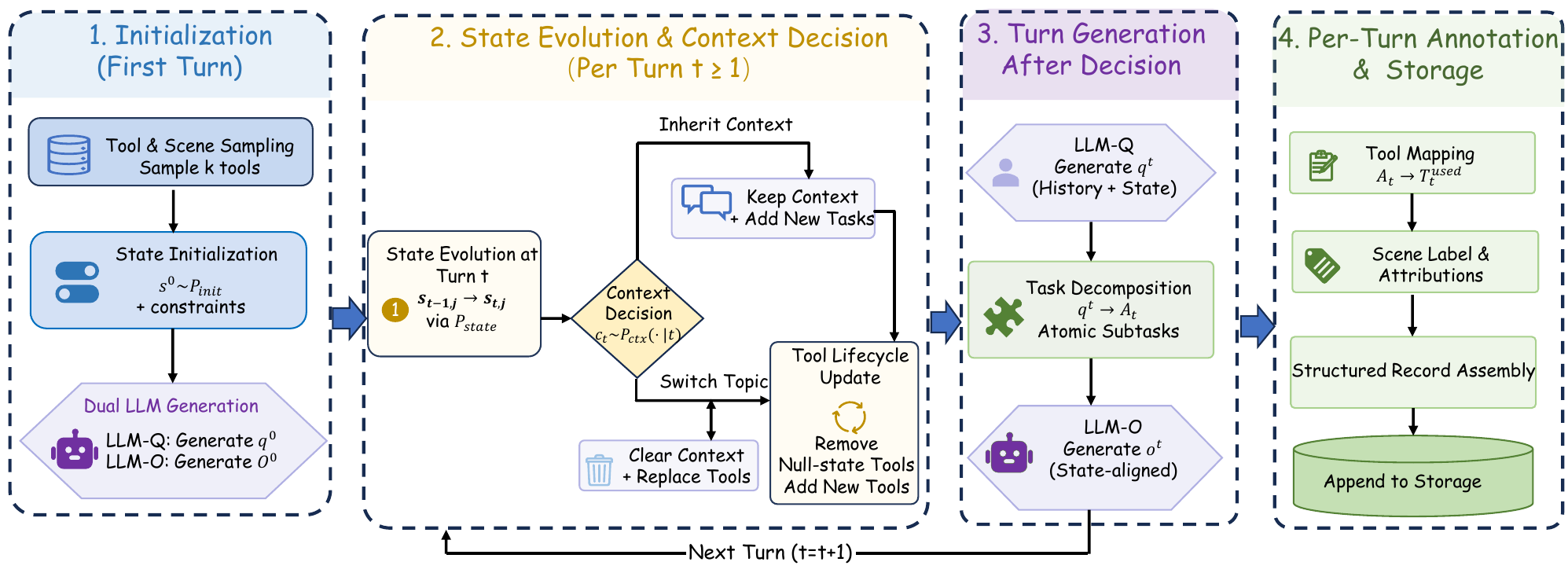}
   \caption{
\textbf{Overview of the MTDTool construction pipeline.}
The data generation follows a four-stage process.
\textbf{(1) Initialization}: Tools and scenarios are sampled to establish initial constraints, and two LLM generators produce the initial user query $q^0$ and assistant output $o^0$.
\textbf{(2) State Evolution \& Context Decision}: At each turn $t\geq 1$, the system decides whether to inherit the previous context or switch to a new topic, and updates the pending intent $p_{t,j}$ and execution state $s_{t,j}$ for each active tool.
\textbf{(3) Turn Generation}: Conditioned on the dialogue history $H_t$, current states, and active tools, the framework generates the user query $q_t$, derives the atomic task sequence $\mathcal{A}_t$, and produces the state-aligned assistant output $o_t$.
\textbf{(4) Per-Turn Annotation \& Storage}: The framework maps the atomic task sequence $\mathcal{A}_t$ to the used tool set $\mathcal{T}^{\mathrm{used}}_t$, assigns scenario labels and attribution information, and stores the structured record.
This pipeline preserves the coherence of multi-turn intent evolution while providing fine-grained annotations for task decomposition and tool retrieval.
}
    \label{fig:data_construction_pipeline}
    \vspace{-13pt}
\end{figure}

\subsection{Dataset Construction via a State-Machine-Driven Framework}
\label{sec:dataset_construction}

To address the evaluation needs for fine-grained task decomposition in mobile multi-turn tool invocation, this paper proposes an automated dialogue generation framework driven by a state machine. This framework models multi-turn interactions as a stochastic decision process driven by state evolution, uniformly characterizing context transitions, task decomposition, and tool execution behaviors. In contrast to traditional construction paradigms that focus solely on input-output pairs, our method explicitly models the dynamic dependencies among dialogue states, user intents, and tool calls, thereby generating high-fidelity multi-turn corpora with fine-grained structural annotations.

\subsubsection{State Space Definition and Dynamic Evolution Mechanism}
\label{subsec:state_space}
As illustrated in Figure~\ref{fig:data_construction_pipeline}, we model the construction of multi-turn dialogues as a state-driven stochastic evolution process. Let $\mathcal{S}$ denote the dialogue state space, which contains 11 canonical interaction behaviors (see Table~\ref{tab:state_definitions}), including task execution states (e.g., success $s_1$, in-progress $s_2$, and failure $s_3$), interaction states (e.g., clarification $s_4$ and selection $s_5$), and intent-management states (e.g., silence $s_6$, cancellation $s_7$, revocation $s_8$, supplementation $s_{10}$, and null $s_{11}$). Given a tool library $\mathcal{T}$, each tool $a_j$ is associated with a semantically compatible state subset
$\mathcal{S}_{a_j}\subseteq\mathcal{S}$., which constrains the feasible interaction states of that tool.
During initialization, each tool's state is sampled from a tool-conditioned prior:
\begin{equation}
    s_{0,i} \sim P_{\mathrm{init}}(\cdot \mid a_i), 
    \quad s_{0,i}\in \mathcal{S}_{a_i},
\end{equation}
where $P_{\mathrm{init}}$ controls the prior frequency of basic interaction phenomena. We further impose lightweight constraints to ensure that the generated dialogues contain sufficient multi-turn interaction signals, such as clarification, continuation, or silence.
Subsequently, the evolution process comprises context decision, intent transition, and state transition. At turn $t$, the framework first samples a context decision
\begin{equation}
    c_t \sim P_{\mathrm{ctx}}(\cdot \mid t), 
    \quad c_t\in\{\mathrm{inherit},\mathrm{switch}\},
\end{equation}
which determines whether the dialogue continues the previous context or switches to a new topic. For each active tool $a_j$, the framework then samples a user-side pending intent and derives the corresponding execution state:
\begin{equation}
    p_{t,j} \sim P_{\mathrm{intent}}(\cdot \mid s_{t-1,j}, c_t),
\end{equation}
\begin{equation}
    s_{t,j} \sim P_{\mathrm{state}}(\cdot \mid p_{t,j}, s_{t-1,j}),
    \quad s_{t,j}\in \mathcal{S}_{a_j}.
\end{equation}
This decoupled formulation separates user intent evolution from assistant execution feedback, making the generation process more controllable. Finally, we apply a lifecycle update to the active tool set: if both the pending intent $p_{t,j}$ and the execution state $s_{t,j}$ are assigned to the null state, tool $a_j$ is removed from the current context.

\subsubsection{Context Control and Task Composition Strategy}
\label{subsec:context_evolution}

To simulate the topic convergence characteristics of real-world dialogues, the framework introduces a context decision variable $c_t \in \{\mathrm{inherit}, \mathrm{switch}\}$ at each evolution node. The probability of switching topics follows an exponential decay function:
\begin{equation}
    P_{\mathrm{switch}}(t) = \max(p_{\min}, p_0 \cdot \alpha^{t-1}),
\end{equation}
where $p_0$ denotes the baseline probability of topic switching in the initial turn, $\alpha$ is the decay coefficient governing the rate of topic convergence, and $p_{\min}$ sets the lower bound for maintaining minimal topic diversity in later dialogue stages. This mechanism ensures high intent variability in early turns while progressively enhancing contextual coherence and stability as the dialogue progresses. At the task structure level, the system supports proportional sampling of three modes: single-tool, multi-tool collaboration, and multi-intent parallelism, while allowing dynamic injection of new tasks during context inheritance. The tool sampling phase combines global random weighting with scene-based knowledge base guidance, utilizing pre-defined or LLM-dynamically-generated scene templates to ensure logical self-consistency of tool combinations within specific domains.

\subsubsection{Dual-Channel Decoupled Generation and Quality Constraints}
\label{subsec:decoupled_generation}

Corresponding to the content synthesis module in the pipeline, the framework employs a dual-channel mechanism to generate user queries $q_t$ (as shown in Appendix~\hyperref[ap:user_prompt]{prompt 1}) and assistant responses $o_t$ (as shown in Appendix~\hyperref[ap:assistant_prompt_example]{prompt 2}) separately. The user-side generator integrates available tool lists, historical state trajectories, and linguistic style features, employing natural language strategies to avoid mechanical concatenation and explicitly completing missing information dimensions in clarification states. The assistant-side generator strictly adheres to state-semantic mapping rules, ensuring precise alignment between response content and current execution states (e.g., success confirmation, failure attribution).

To guarantee corpus quality, we embed a structured hard-constraint layer into the generation process, including: (i) the one-tool-one-request atomicity principle; (ii) colloquial conversion of terminology; (iii) normalization of temporal expressions; and (iv) zero-response specifications for silence states. These constraints act as quality filters throughout the generation process, ensuring that the output data possesses both linguistic naturalness and logical rigor.

\subsubsection{Fine-Grained Annotation and Evaluation Support}
\label{subsec:annotation_evaluation}

At the output end of the pipeline, an independent task decomposition module maps user requests containing multiple intents into an atomic task sequence $\mathcal{A}_t$ and establishes semantic associations with the candidate tool set $\mathcal{T}_{\mathrm{cand},t}$:
\begin{equation}
  q_t, H_t \rightarrow \mathcal{A}_t \rightarrow \mathcal{T}_{\mathrm{cand},t}.
\end{equation}
This process constructs a three-level annotation structure (\textit{Dialogue Context}--\textit{Atomic Tasks}--\textit{Candidate Tools}), which not only supports standard ranking metrics but also precisely locates model bottlenecks such as context misunderstanding, task decomposition errors, or tool retrieval failures. The final dataset includes complete fields such as Dialogue ID, Turn Index, Scene Label, Interaction Content, Execution States, Intent Trajectories, and Decomposition Results, forming the foundation of a high-quality structured corpus. A full example of the multi-turn dialogue generation process is shown in Appendix~\hyperref[ap:full_prompt]{3}.

\begin{figure}[hb]
    \centering
    \includegraphics[width=\textwidth]{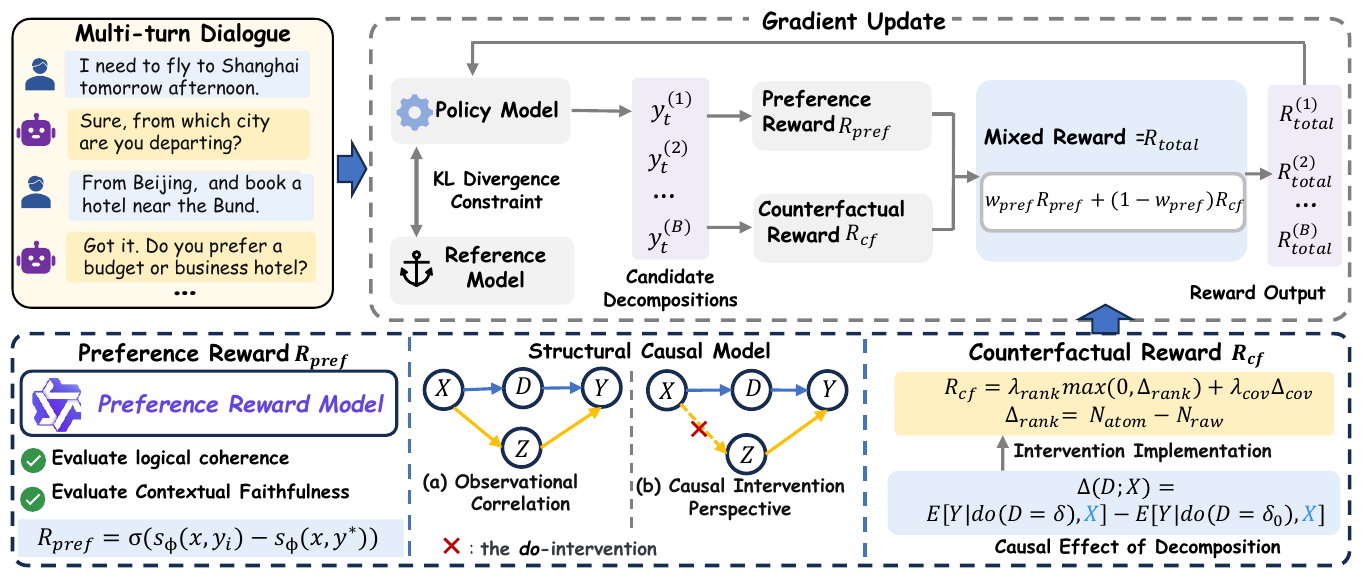}
    \caption{Overview of the PCTD.
     }
    \label{fig:algorithm_framework}
    \vspace{-13pt}
\end{figure}

\subsection{Joint Reward Mechanism}
In the scenario of contextual task atomization, the efficacy of model-generated decomposition results should not be determined solely by superficial textual rationality, but rather by their actual contribution to downstream tool retrieval performance and the semantic quality of the decomposition structure itself. To this end, we propose a joint reward mechanism that integrates counterfactual~\citep{verma2024counterfactual,guo2025counterfactual} causal feedback with preference alignment, guiding the policy model to learn optimal semantic decomposition paths.

\subsubsection{Counterfactual Baseline Construction and Reward Mechanism}
To characterize the true contribution of task decomposition to tool retrieval, we construct a Structural Causal Model (SCM), as illustrated in Figure~\ref{fig:algorithm_framework}. Specifically, $X$ denotes the multi-turn dialogue input, $D$ denotes the task decomposition strategy, $Y$ denotes the retrieval ranking quality, and $Z$ represents confounding factors such as high-frequency keywords, tool priors, and model parametric memory. Traditional methods directly optimize outcome-oriented metrics such as Recall or NDCG, which are susceptible to the spurious path $X \rightarrow Z \rightarrow Y$. As a result, retrieval gains caused by raw-query keywords or tool priors may be incorrectly attributed to task decomposition, inducing shallow strategies such as repetitive decomposition and keyword stacking.
To mitigate this attribution bias, we introduce a paired counterfactual comparison. Given the same input, tool library, retriever, and ground-truth tool set, we jointly evaluate the decomposition branch $d$ and the non-decomposition baseline $d_0$. Since the two branches share the same retrieval environment, confounding factors such as keyword priors, tool distributions, and parametric memory are held fixed and subtracted out through differencing. Therefore, the residual performance difference can be used as a controlled estimate of the marginal contribution of task decomposition.
Formally, we treat task decomposition as an intervention on the retrieval process and define the marginal causal contribution of decomposition as
\begin{equation}
    \Delta(D; X) =
    \mathbb{E}\left[Y \mid do(D=\delta), X\right]
    -
    \mathbb{E}\left[Y \mid do(D=\delta_0), X\right],
\end{equation}
where $\delta$ is the atomized decomposition strategy generated by the policy model, and $\delta_0$ is the non-decomposition baseline, i.e., directly retrieving tools using the raw context $X$. Under the paired-counterfactual assumption that the nuisance factors $Z$ are shared between the two branches conditioned on the same input $X$, this difference provides a controlled estimate of the additional retrieval benefit brought by task decomposition.

In practice, we instantiate the potential outcome $Y$ with the retrieval ranking metric NDCG@K. 
Given the multi-turn dialogue context $x=(H,q)$ and the ground-truth tool set $\mathcal{G}_t$, we first evaluate the non-decomposition branch by directly retrieving tools with the raw context:
\begin{equation}
    \mathcal{L}_{\mathrm{raw}}^K = R_K(x),\quad
    N_{\mathrm{raw}} = \mathrm{NDCG@K}
    \left(\mathcal{L}_{\mathrm{raw}}^K, \mathcal{G}_t\right),
\end{equation}
where $R_K(\cdot)$ denotes the top-$K$ retriever over the tool library $\mathcal{T}$.
For the decomposition branch, a candidate decomposition $y_{t,i}$ is parsed into an ordered sequence of atomic subtasks:
\begin{equation}
 \mathcal{A}_{t,i} = \mathrm{Parse}(y_{t,i})
= (\tau_{t,i,1}, \tau_{t,i,2}, \ldots, \tau_{t,i,m_{t,i}}).
\end{equation}
Each subtask is used as an independent retrieval query. The retrieved tool lists are then merged, deduplicated, and truncated to obtain the decomposition-aware ranked list:
\begin{equation}
    \mathcal{L}_{\mathrm{atom}}^K
    =
    \operatorname{TopK}
    \left(
    \operatorname{Dedup}
    \left(
    \operatorname{Merge}
    \left(
   R_K(\tau_{t,i,1}), \ldots, R_K(\tau_{t,i,m_{t,i}})
    \right)
    \right)
    \right).
\end{equation}
The retrieval quality of the decomposition branch is computed as:
\begin{equation}
    N_{\mathrm{atom}}
    =
    \mathrm{NDCG@K}
    \left(\mathcal{L}_{\mathrm{atom}}^K, \mathcal{G}_t\right).
\end{equation}
The paired comparison defines the ranking-level counterfactual gain as:
\begin{equation}
    \Delta_{\mathrm{rank}} = N_{\mathrm{atom}} - N_{\mathrm{raw}} .
\end{equation}
To capture whether decomposition improves complete tool coverage, we further define:
\begin{equation}
    \Delta_{\mathrm{cov}}
    =
    \mathbb{I}\left[\mathcal{G}_t\subseteq \mathcal{L}_{\mathrm{atom}}^K\right]
    -
    \mathbb{I}\left[\mathcal{G}_t\subseteq \mathcal{L}_{\mathrm{raw}}^K\right],
\end{equation}
where $\mathbb{I}[\cdot]$ is the indicator function. The counterfactual retrieval-gain reward is then formulated as:
\begin{equation}
    R_{\mathrm{cf}}
    =
    \lambda_{\mathrm{rank}}\max(0,\Delta_{\mathrm{rank}})
    +
    \lambda_{\mathrm{cov}}\Delta_{\mathrm{cov}} .
\end{equation}
Here, the first term rewards decompositions that improve NDCG@K over the raw-query baseline, while the second term measures the gain or loss in complete tool coverage. Since both branches share the same input, retriever, tool library, and ground-truth tool set, this reward reduces attribution bias from raw-query keywords and tool priors, encouraging decompositions that provide additional retrieval benefits.

\subsection{Preference Reward Modeling}
To compensate for the semantic blind spots of pure retrieval metrics, we introduce an independent preference reward $R_{\mathrm{pref}}$. This reward is based on a specially trained Process Reward Model (PRM) that evaluates decomposition results across five dimensions: completeness, accuracy, coreference resolution, standardized expression, and contextual consistency (detailed descriptions and penalty rules for each dimension are provided in Appendix Table~\ref{tab:prm_dimensions}).

Let $s_\phi(x,y)$ denote the raw quality logit assigned by the PRM to a decomposition $y$ under context $x$, and let $y^\star$ denote the human-annotated reference decomposition. For each candidate decomposition $y_{t,i}$, we construct a reference-anchored preference pair $(y_i,y^\star)$ and define the preference reward as:
\begin{equation}
   R_{\mathrm{pref}}(x_t,y_{t,i})
=
\sigma\left(
s_\phi(x_t,y_{t,i})-s_\phi(x_t,y^\star_t)
\right),
\end{equation}
where $\sigma(\cdot)$ is the Sigmoid function. This Bradley--Terry-style reward measures the preference strength of the candidate decomposition relative to the expert reference. Since $y^\star$ is fixed under the same context, this design provides a stable quality anchor and is further converted into a relative advantage through group-wise normalization in GRPO, thereby offering dense structural supervision for logical coherence, contextual faithfulness, completeness, and atomicity. During GRPO optimization, we combine the counterfactual retrieval reward and the preference reward into a unified training signal:
\begin{equation}
    R
    =
    w_{\mathrm{pref}} R_{\mathrm{pref}}
    +
    (1-w_{\mathrm{pref}}) R_{\mathrm{cf}},
\end{equation}
where $w_{\mathrm{pref}}\in[0,1]$ controls the trade-off between structural decomposition quality and counterfactual retrieval gain. The preference reward $R_{\mathrm{pref}}$ encourages the policy to generate logically coherent, context-faithful, and atomic decompositions, while the counterfactual reward $R_{\mathrm{cf}}$ promotes decompositions that improve tool ranking and coverage beyond the raw-query baseline. By balancing these complementary signals, the policy is optimized via Eq.~\ref{eq: grpo} to produce decompositions that are both structurally sound and effective for retrieval. The complete algorithm is detailed in Appendix~\ref{alg:cf_gdpo}.


\section{Experiments}

\subsection{Evaluation Benchmarks}
We evaluate PCTD on two benchmarks: ToolRet~\citep{shi2025retrieval} and MTDTool. ToolRet aggregates 35 widely-used tool-calling datasets spanning three domains (Web, Code, and Custom), with a total corpus of 44K tools. MTDTool is a dataset specifically constructed for complex multi-turn mobile interaction scenarios, covering 237 vertical-domain tools across diverse mobile contexts. The training and test splits of ToolRet strictly follow the setup in ToolQP: we sample 10k instances from Web-domain data and synthesize tool-calling trajectories using Qwen3.6-Plus to form the training set; among these, 3.1k Web samples serve as the In-Domain test set, with all involved tools exposed during training, while the remaining test data in ToolRet constitute the Out-of-Domain test set. For MTDTool, we partition the test set into In-Domain and OOD subsets based on tool vertical domains. We use NDCG@K and Completeness@K $(1\,[R@K = 1])$ as evaluation metrics, denoted as N@K and C@K in this paper.
Detailed specifications are provided in the Appendix~\ref{ap: detail_dataset_info}.

\subsection{Baselines}
To comprehensively evaluate task decomposition and tool retrieval performance, we compare against three representative categories of baselines: (1) closed-source LLMs (ChatGPT-5.0\footnote{https://chat.openai.com/}, Qwen3-series~\citep{yang2025qwen3}, MiniMax-M2.5\footnote{https://minimaxi.com/news/minimax-m25}, and DeepSeek-V4-Pro\footnote{https://www.DeepSeek.com/}), which serve to establish the upper bound of general-purpose models in zero-shot or few-shot decomposition scenarios; (2) prompting-based methods (Q2E~\citep{wang2023query2doc}, ReInvoke~\citep{chen2024re}, ToolReAGt~\citep{brauns2025toolreagt}, and PLUTO~\citep{huang2024pluto}), which optimize decomposition strategies via explicit chain-of-thought or structured retrieval augmentation, representing lightweight paradigms without parameter updates; and (3) the RL-based end-to-end optimization method ToolQP~\citep{fang2026beyond}, which incorporates downstream tool retrieval metrics into mixed reward signals to jointly optimize decomposition and retrieval processes. Detailed descriptions of all baselines are provided in the Appendix~\ref{baseline}. Detailed implementation settings are provided in the Appendix~\ref{ap:implementation_details}.

\begin{table*}[ht]
\centering
\small
\setlength{\tabcolsep}{4pt}
\renewcommand{\arraystretch}{1.1}
\caption{Overall performance comparison of PCTD and baselines across two benchmarks. Best and second-best results are bolded and underlined, respectively.}
\label{tab: main_results}
\begin{tabular}{lcccccccccccc}
\toprule
\multirow{3}{*}{\textbf{Methods}}
& \multicolumn{4}{c}{\textbf{In-Domain}}
& \multicolumn{4}{c}{\textbf{Out-of-Domain}}
& \multicolumn{2}{c}{\textbf{Average}} \\

\cmidrule(lr){2-5}
\cmidrule(lr){6-9}
\cmidrule(lr){10-11}

& \multicolumn{2}{c}{ToolRet }
& \multicolumn{2}{c}{MTDTool}
& \multicolumn{2}{c}{ToolRet}
& \multicolumn{2}{c}{MTDTool}
& \multicolumn{2}{c}{Overall} \\

\cmidrule(lr){2-3}
\cmidrule(lr){4-5}
\cmidrule(lr){6-7}
\cmidrule(lr){8-9}
\cmidrule(lr){10-11}

& N@10 & C@10
& N@10 & C@10
& N@10 & C@10
& N@10 & C@10
& N@10 & C@10 \\

\midrule
\multicolumn{11}{l}{\textit{\textbf{Base Retriever}}} \\
Qwen3-4B-embedding  &56.82	&59.32	&68.35	&61.79	&36.98	&38.10	&50.57	&42.22	&53.18 	&50.36 \\

\midrule
\multicolumn{11}{l}{\textit{\textbf{Closed-Source LLMs}}} \\
ChatGPT-5.0 &60.30 &64.74 &82.81 &82.50 &37.96 &39.53 &73.22 &74.24 &63.57 &65.25
 \\
DeepSeek-V4-Pro &58.23 &61.94 &83.06 &78.70 &32.85 &33.12 &73.26 &69.22 &61.85 &60.75\\
MiniMax-M2.5 &60.32 &63.65 &84.11 &79.18 &32.00 &32.60 &74.51 &70.18 &62.74 &61.40 \\
\midrule
\multicolumn{11}{l}{\textit{\textbf{Prompting Methods (Qwen3-4B)}}} \\
Q2E &53.80 &56.48 &70.36 &65.43 &29.57 &29.88 &54.84 &46.39 &52.14 &49.55\\
ReInvoke &59.55 &63.55 &85.68 &\underline{80.26} &34.57 &34.97 &75.08 &\underline{69.37} &63.72 &\underline{62.04} \\
ToolReAGt &51.79 &57.48 &77.09 &68.81 &29.26 &34.24 &72.54 &64.58 &57.67 &56.28\\
PLUTO  &\underline{59.96} & \underline{65.82} &77.69 &56.29 &34.40 &34.24 &\underline{77.29} &58.11 &62.34 &53.62\\

\hdashline
\multicolumn{11}{l}{\textit{\textbf{RL-Based Methods (Qwen3-4B)}}} \\
ToolQP &56.27 &62.58 &\underline{86.72} &76.99 &\underline{37.67} & \underline{39.93} &74.60 &68.28 &\underline{63.82} &61.95 \\
\rowcolor{blue!8} \textbf{Ours} &\textbf{64.71} &\textbf{68.71} &\textbf{89.17} &\textbf{84.11} &\textbf{39.27} &\textbf{41.11} &\textbf{78.82} &\textbf{74.87} &\textbf{67.99} &\textbf{67.20}\\
$\Delta$ &4.75 &2.89 &2.45 &3.85 &1.60 &1.18 &1.53 &5.50 &4.17 &5.16 \\
\midrule
\multicolumn{11}{l}{\textit{\textbf{Prompting Methods (Qwen3-8B)}}} \\
Q2E &55.71 &57.74 &70.43 &66.03 &36.47 &37.80 &54.66 &46.32 &54.32 &51.97 \\
ReInvoke &\underline{57.84} &61.94 &84.74 &80.01 &34.14 &34.51 &74.93 &70.79 &62.91 &61.81
\\
ToolReAGt &52.24 &58.52 &81.20 &74.57 &31.94 &35.08 &76.98 &70.12 &60.59 &59.57
\\
PLUTO &56.73 & \underline{64.25} &76.91 &61.73 &33.16 &35.04 &75.93 &60.36 &60.68 &55.35\\
\hdashline
\multicolumn{11}{l}{\textit{\textbf{RL-Based Methods (Qwen3-8B)}}} \\
ToolQP &56.47 &63.23 &\underline{89.72} &\underline{83.61} &\underline{37.77} &\underline{39.36} &\underline{77.92} &\underline{74.33} &\underline{65.47} &\underline{65.13} \\
\rowcolor{blue!8} \textbf{Ours} & \textbf{64.79} &\textbf{68.71} &\textbf{91.19} &\textbf{87.08} &\textbf{39.82} &\textbf{41.50} &\textbf{82.74} &\textbf{79.77} &\textbf{69.64} &\textbf{69.27}

\\
$\Delta$ &6.95 &4.46 &1.47 &3.47 &2.05 &2.14 &4.82 &5.44 &4.17 &4.14 \\

\bottomrule
\end{tabular}
\end{table*}

\subsection{Main Results}
Table~\ref{tab: main_results} presents a comprehensive performance comparison between PCTD and various baseline methods on the ToolRet and MTDTool benchmarks, yielding the following key findings. 
\textbf{First}, to address the OOD generalization failure of existing RL methods caused by their reliance on surface keyword co-occurrences, PCTD demonstrates significant robustness by cutting off spurious correlations via counterfactual rewards. Under the OOD setting of MTDTool with Qwen3-8B as the base model, the N@10 of PCTD reaches 82.74, an improvement of 4.82 over the second-best RL baseline ToolQP (77.92). It also achieves an N@10 of 39.82 on the ToolRet benchmark under the OOD setting, outperforming ToolQP (37.77). This directly proves that the model has genuinely learned the causal logic for guiding retrieval rather than merely memorizing the tool distribution of the training set, thereby effectively mitigating performance collapse in scenarios involving unseen tools.
\textbf{Second}, addressing repetitive decomposition and reward hacking caused by the lack of structural constraints (i.e., normative constraints on task decomposition results across logical, semantic, and execution dimensions), PCTD achieves a substantial leap in decomposition quality while maintaining high retrieval accuracy. In the In-Domain scenario of MTDTool, the C@10 of PCTD (Qwen3-8B) reaches 87.08, significantly outperforming ToolQP (83.61) by 3.47. Combined with the negative correlation between the repetition rate and performance in Figure~\ref{fig: motivation} (where performance increases to 91.2 as the repetition rate drops from 5.2\% to 0.7\%), this improvement verifies that the $R_{\text{pref}}$ effectively eliminates reward hacking behaviors by suppressing the generation of redundant sub-tasks, enabling the model to output more precise and high-quality decomposition strategies.

Ultimately, PCTD (Qwen3-8B) achieves the best performance across all evaluation settings (In-Domain, Out-of-Domain, and Average) on both benchmarks, with average N@10 and C@10 reaching 69.64 and 69.27, respectively. It comprehensively surpasses prompt engineering methods such as PLUTO and ToolReAGt, as well as RL methods like ToolQP, establishing the effectiveness of the causal gain-structural constraint joint optimization paradigm in resolving the fundamental limitations of task decomposition. Additionally, we observe that some prompting methods (e.g., Q2E and ReInvoke) perform slightly worse on the 8B model than on the 4B model. This suggests that fixed prompting strategies are sensitive to model scale, as larger models may suffer from over-reasoning or instruction-following deviations, failing to fully benefit from the increased parameter size. In Table~\ref{ap: retrievers}, we further analyze the performance variations of the PCTD method when using different retrieval models, with detailed analysis results provided in Appendix~\ref{ap: pctd_with_retrievers}.

\begin{wraptable}[15]{r}{0.56\textwidth}
\vspace{-2pt}
\centering
\small
\setlength{\tabcolsep}{2pt}
\renewcommand{\arraystretch}{1.1}
\caption{Ablation study on PCTD.}

\begin{tabular}{lcccc}
\hline
\multirow{2}{*}{\textbf{Methods}} & \multicolumn{2}{c}{In-Domain} & \multicolumn{2}{c}{Out-of-Domain} \\
\cmidrule(lr){2-3}
\cmidrule(lr){4-5}
 & N@10 & C@10 & N@10 & C@10 \\
\hline
Base Retriever & 68.35 & 61.79 & 50.57 & 42.22 \\
\hline
\multicolumn{5}{c}{\textit{\textbf{Prompting vs SFT vs RLVR}}} \\
\hdashline
+ Prompt &81.27 &79.88 &71.44 &70.52 \\
+ SFT & 85.35 & 79.43 & 75.23 & 71.08 \\
\rowcolor{blue!8} + RLVR & \textbf{91.19} & \textbf{87.08} & \textbf{82.74} & \textbf{79.77} \\
\hline
\multicolumn{5}{c}{\textit{\textbf{RL-reward Ablation 
Study}}} \\
\hdashline
w/o Preference Reward & 90.65 & 86.18 & 78.59 & 75.79 \\
w/o Counterfactual Reward & 89.04 & 84.37 & 77.29 & 73.99 \\
\rowcolor{blue!8} PCTD & \textbf{91.19} & \textbf{87.08} & \textbf{82.74} & \textbf{79.77} \\
\hline
\end{tabular}
\end{wraptable}

\subsection{Ablation Studies}
To validate the effectiveness of each component within PCTD, we conducted systematic ablation studies on the MTDTool benchmark. \textbf{First}, in the comparison of training paradigms, while the Prompting approach outperforms the base retriever without decomposition, it remains constrained by the model's inherent knowledge, resulting in OOD performance (N@10 of 71.44) that significantly lags behind parameterized methods. Although Supervised Fine-Tuning (SFT) improves In-Domain (ID) performance, its OOD generalization hits a bottleneck (N@10 of merely 75.23) due to overfitting to surface-level patterns. In contrast, the reinforcement learning-based PCTD achieves a significant breakthrough via end-to-end optimization, reaching an OOD N@10 of 82.74 (a 7.51-point improvement over SFT). This fully demonstrates the advantage of RL in capturing genuine causal logic.
\textbf{Second}, the ablation results of the reward mechanism validate the necessity of our dual-track design. Removing the preference reward impairs the model's robustness in OOD scenarios due to the lack of structural constraints on logical coherence and atomicity, causing a 4.15-point drop in N@10. The performance degradation is even more severe when the counterfactual reward is removed, with the OOD N@10 dropping significantly by 5.45 points. This indicates that without the guidance of marginal causal gain, the model degenerates into shallow matching that relies on keyword co-occurrence. Ultimately, the complete PCTD achieves optimal performance in both retrieval accuracy and generalization capability by synergizing the functional correction of counterfactual rewards with the structural constraints of preference rewards.

\begin{wraptable}[17]{r}{0.7\textwidth}
\vspace{-2pt}
\centering
\small
\setlength{\tabcolsep}{1pt}
\renewcommand{\arraystretch}{1.1}
\caption{the scenario-level difficulty and discriminative capability of MTDTool.}
\label{tab:dataset_quality}

\begin{tabular}{clcccc}
\hline
\multirow{2}{*}{\textbf{Scenario}} & \multirow{2}{*}{\textbf{Type}} & \multicolumn{2}{c}{In-Domain} & \multicolumn{2}{c}{Out-of-Domain} \\
\cmidrule(lr){3-4}
\cmidrule(lr){5-6}
& & N@10 & C@10 & N@10 & C@10 \\
\hline

\multirow{3}{*}{Single-turn}
& Single Tool & 95.48 & 100.00 & 81.95 & 95.36 \\
& Same-category Multi-tool & 93.99 & 95.42 & 77.77 & 63.78 \\
& Cross-category Multi-tool & 86.02 & 73.27 & 74.41 & 55.48 \\
\hline

\multirow{7}{*}{Multi-turn}
& Domain Switch & 91.78 & 95.89 & 81.27 & 93.08 \\
& Multi-turn Continuation & 81.29 & 83.83 & 69.31 & 74.79 \\
& Intent Selection & 95.09 & 99.29 & 82.88 & 94.13 \\
& User Adds New Task & 94.31 & 98.58 & 84.64 & 89.51 \\
& Unfinished Task Continuation & 90.71 & 87.28 & 73.00 & 64.64 \\
& Abnormal Scenario & 90.20 & 92.68 & 77.89 & 81.65 \\
& State Hybrid & 88.48 & 70.35 & 76.79 & 53.36 \\
\hline

\end{tabular}
\end{wraptable}

\subsection{Dataset Evaluation}
We categorize MTDTool into single-turn and multi-turn dialogues with 10 fine-grained scenario labels, and evaluate the performance under both In-Domain and OOD settings. The experimental results are shown in Table~\ref{tab:dataset_quality}.
The scenario-level results further reveal clear difficulty gradients. In single-turn settings, performance decreases as tool composition becomes more complex: Single Tool achieves the best In-Domain N@10 of 95.48, while Cross-category Multi-tool drops to 86.02 In-Domain and 74.41 OOD, with a notable coverage decline. In multi-turn settings, structured scenarios such as Intent Selection and User Adds New Task remain relatively easy, reaching 95.09 and 94.31 In-Domain N@10, respectively. In contrast, scenarios requiring stronger state tracking, such as Multi-turn Continuation and State Hybrid, show larger degradation, especially in complete coverage, where State Hybrid drops to 53.36 C@10 under OOD.
These results suggest that MTDTool provides a fine-grained and discriminative benchmark for evaluating task decomposition and tool retrieval across tool composition, domain shift, and multi-turn state evolution. We further present an evaluation of decomposition quality in Appendix~\ref{ap: GSB_evalution} using the General Scoring Benchmark (GSB) based on the LLM-as-Judge paradigm.

\begin{figure*}[t]
\centering
\begin{subfigure}{0.24\textwidth}
    \centering
    \includegraphics[width=\linewidth]{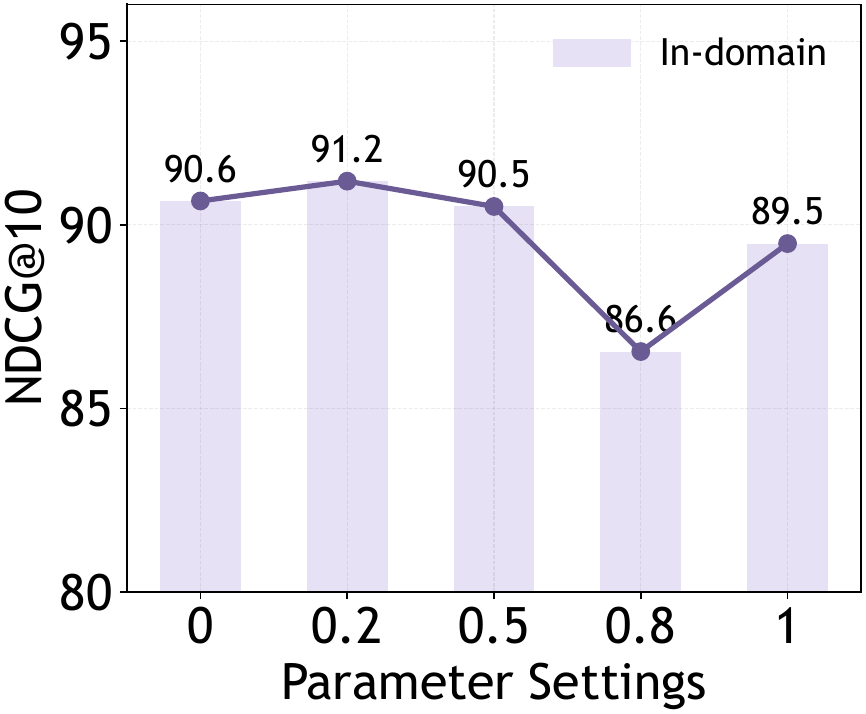}
    \caption{In-Domain}
\end{subfigure}
\begin{subfigure}{0.24\textwidth}
    \centering
    \includegraphics[width=\linewidth]{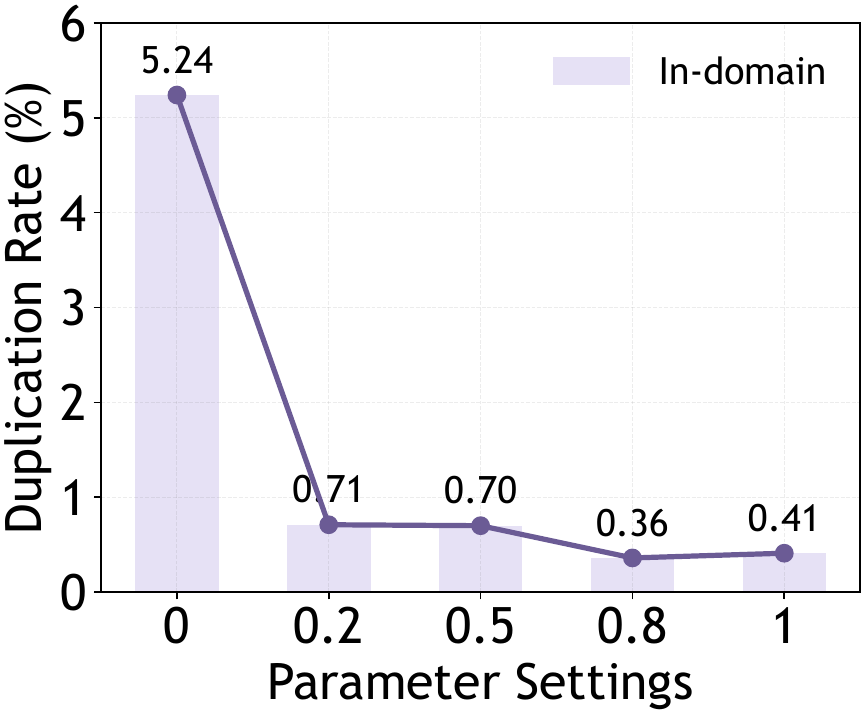}
     \caption{In-Domain}
\end{subfigure}
\begin{subfigure}{0.24\textwidth}
    \centering
    \includegraphics[width=\linewidth]{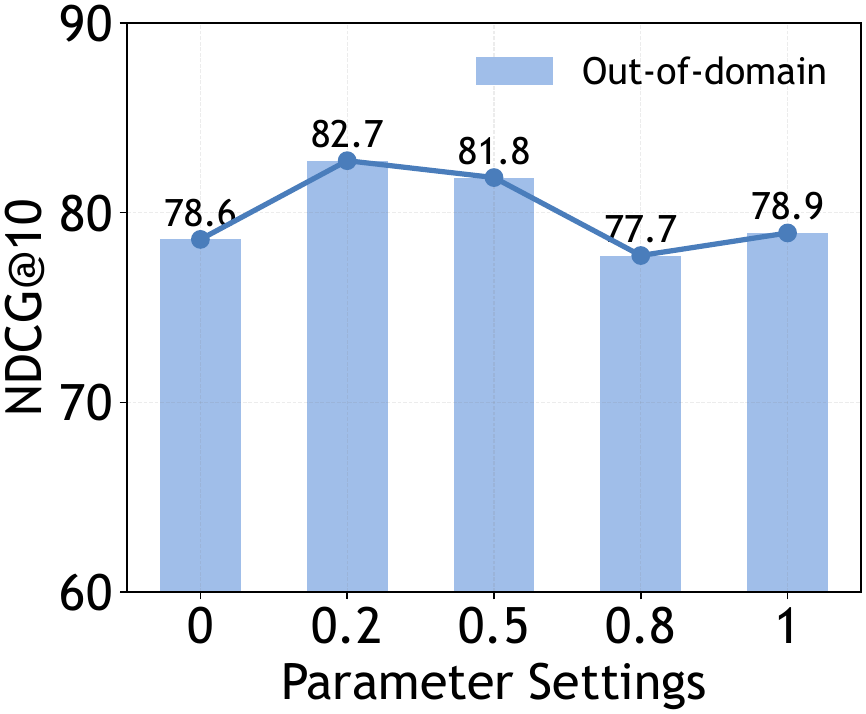}
     \caption{Out-of-Domain}
\end{subfigure}
\begin{subfigure}{0.24\textwidth}
    \centering
    \includegraphics[width=\linewidth]{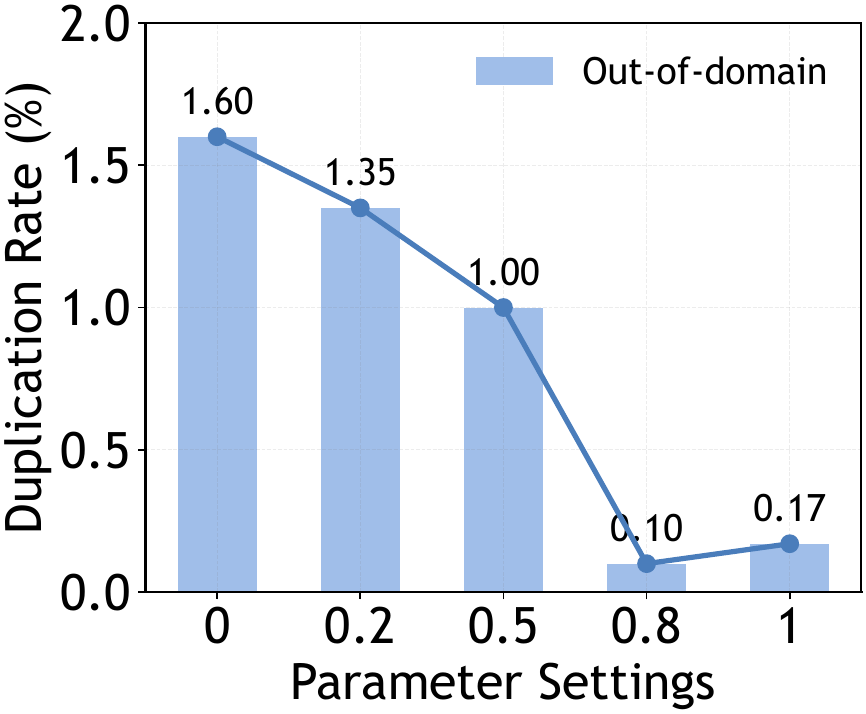}
     \caption{Out-of-Domain}
\end{subfigure}
\caption{Hyperparameter analysis under different $w_{pref}$ settings.}
\label{fig:hyperparameter_analysis}
\end{figure*}
\begin{figure}[hb]
\centering

\begin{subfigure}{0.24\textwidth}
\centering
\includegraphics[width=\linewidth]{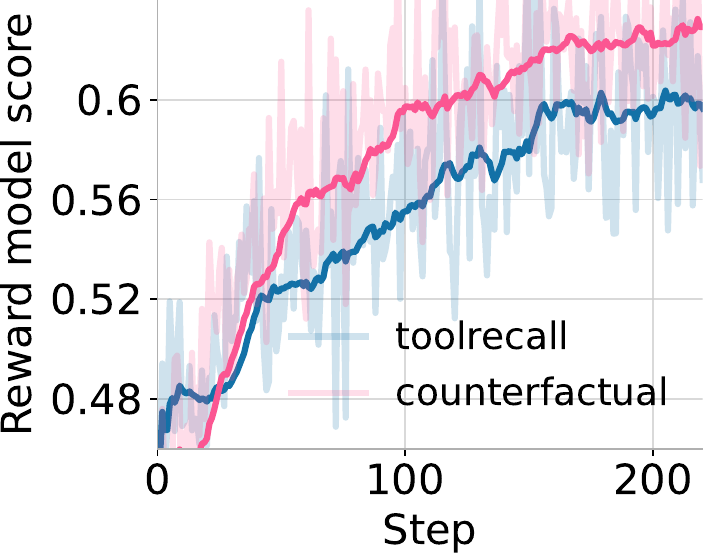}
\caption{}
\end{subfigure}
\begin{subfigure}{0.24\textwidth}
\centering
\includegraphics[width=\linewidth]{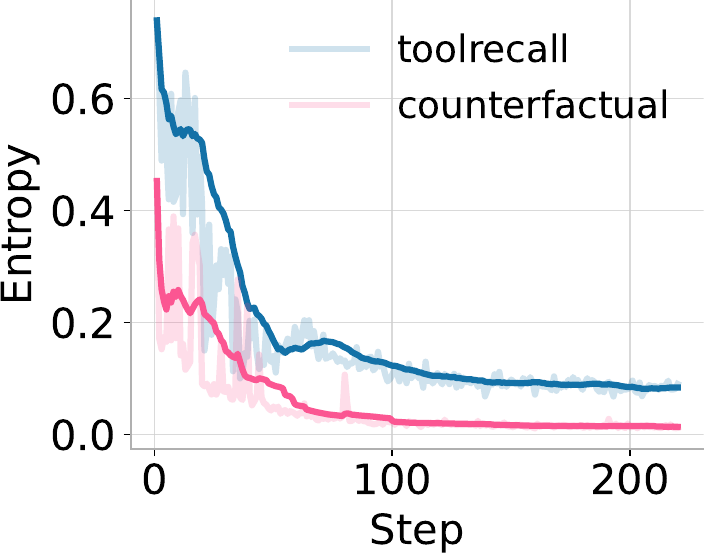}
\caption{}
\end{subfigure}
\begin{subfigure}{0.24\textwidth}
\centering
\includegraphics[width=\linewidth]{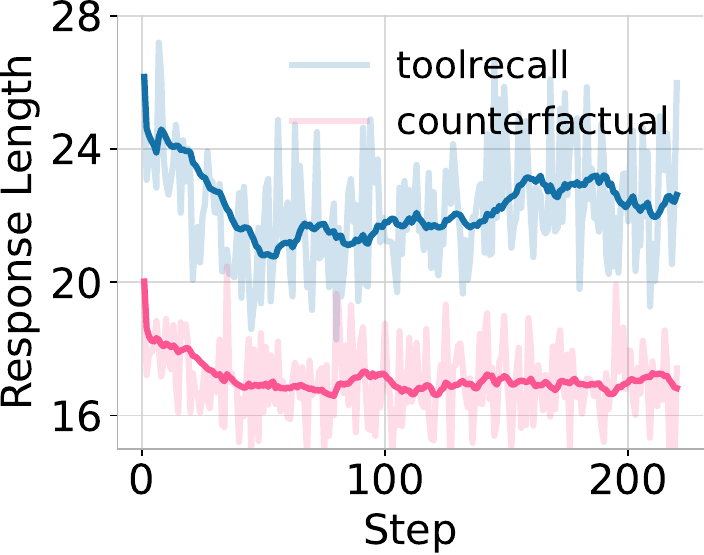}
\caption{}
\end{subfigure}
\begin{subfigure}{0.24\textwidth}
\centering
\includegraphics[width=\linewidth]{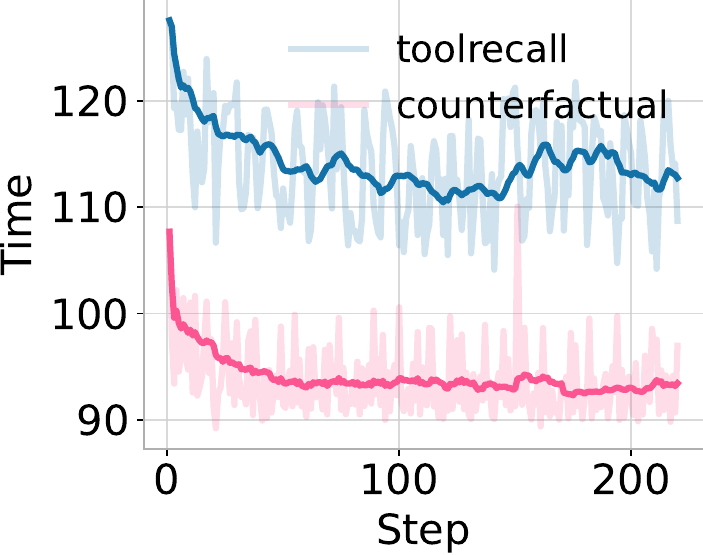}
\caption{}
\end{subfigure}

\caption{Training curves of different metrics.}
\label{fig:training_curves}

\end{figure}
\subsection{Deeper Analysis}

\textbf{When might PCTD fail?} We further analyze that improper reward weight settings can lead to the failure of PCTD. As shown in Figure~\ref{fig:hyperparameter_analysis} (a)-(c), when the preference reward weight $w_{pref}=0$, the model relies solely on counterfactual signals for optimization. Although it maintains high retrieval accuracy In-Domain (N@10 = 90.6), the duplication  rate surges to 5.24\%, and OOD performance drops sharply to 78.6. This confirms that pure causal optimization without preference constraints still induces reward hacking, resulting in generalization failure. Conversely, an excessively high $w_{pref}$ may slightly sacrifice retrieval accuracy due to over-regularization, despite suppressing redundancy. Therefore, the risk of PCTD failure primarily stems from extreme imbalances in reward weights; only by maintaining a dynamic balance between causal gain and structural constraints can dual optimality in retrieval accuracy and generalization robustness be achieved. Detailed hyperparameter analysis is provided in the Appendix~\ref{ap: when_fails}.

We further analyze the training dynamics of various metrics, comparing the existing reinforcement learning baseline relying on traditional tool retrieval rewards with our PCTD framework, as illustrated in Figure~\ref{fig:training_curves}. Specifically, as shown in Figure~\ref{fig:training_curves}(a), the reward model score of PCTD rises more rapidly and reaches a final value of 0.62, significantly outperforming the baseline value of 0.59. This indicates that the preference reward $R_{\text{pref}}$ effectively guides the model to generate high-quality decomposition results that are more logically coherent, highly atomic, and better aligned with human references, thereby compensating for the lack of structural supervision in existing methods. In terms of policy exploration, as depicted in Figure~\ref{fig:training_curves}(b), the policy entropy of PCTD rapidly decreases and stabilizes at a minimal level of 0.05, substantially lower than that of the baseline. This confirms that the counterfactual reward $R_{\text{cf}}$ provides clear causal gradient signals, enabling the model to quickly converge to the optimal decomposition strategy and reducing ineffective exploration in the surface-level lexical space. More crucially, as demonstrated in Figure~\ref{fig:training_curves}(c), the response length of PCTD is significantly shortened to approximately 17 tokens, whereas the baseline remains at 22 tokens. This directly verifies that PCTD successfully suppresses reward hacking behaviors, such as repetitive decomposition common in existing methods, through structural constraints, forcing the model to output refined sub-tasks. Ultimately, benefiting from the streamlined generation sequence and faster convergence, as evidenced in Figure~\ref{fig:training_curves}(d), the per-step training time of PCTD drops to 92 milliseconds, outperforming the 110 milliseconds of the baseline. This demonstrates that while resolving generalization failures and structural redundancies, the proposed method not only avoids additional computational overhead but also achieves an optimization in overall training efficiency.

\section{Related Work}
\subsection{Planning and Tool Retrieval in Agents}
Research on agent planning and tool retrieval has evolved along two primary paradigms~\citep{huang2024understanding, yehudai2025survey}: (1) Prompt-based and In-Context Learning approaches, grounded in the Reasoning-Acting-Reflecting cycle established by ReAct~\citep{yao2022react} and ToolReAct~\citep{brauns2025toolreagt}. Subsequent works (e.g., Toolformer~\citep{schick2023toolformer}, Gorilla~\citep{patil2024gorilla}, and the ToolLLM~\citep{qin2024toolllm} series) have achieved effective tool selection within fixed inventories by explicitly generating chains of thought, sub-goals, or API invocation sequences; and (2) Joint optimization via Retrieval-Augmented Generation (RAG)~\citep{zhao2026retrieval} and Reinforcement Learning~\citep{liu2026llm}, designed to address challenges in large-scale open tool libraries. These methods typically decompose complex instructions into structured sub-queries (e.g., DAG-based atomic planning in Plan-RAG~\citep{verma2024plan} or iterative query planning in TOOLQP~\citep{fang2026beyond}) and refine policies using synthetic trajectories and RLVR. Recent studies (e.g., ToolRL~\citep{qian2026toolrl}, RLTR~\citep{li2025encouraging}, GiGPO~\citep{feng2026group}) further incorporate GRPO~\citep{guo2025DeepSeek}, fine-grained tool-use rewards, or hierarchical advantage estimation to enhance end-to-end performance in long-horizon planning and multi-step invocation.
However, existing paradigms directly use retrieval metrics as rewards, which easily induces reward hacking by relying on spurious correlations between shallow lexical features of decompositions and retrieval outcomes, leading to OOD generalization failure and repetitive decomposition. To address this, we propose the PCTD framework, which cuts off spurious correlations via counterfactual rewards and synergizes preference rewards to enforce fine-grained structural constraints, achieving joint optimization of task planning and tool retrieval in both causal gain and structural quality.

\subsection{Benchmarks for Tool Retrieval}
Existing benchmarks for agent task decomposition and tool retrieval fall into two primary categories:  tool-interaction benchmarks (e.g., API-Bank~\citep{li2023api}, ToolBench~\citep{qin2024toolllm}, MINT-Bench~\citep{wang2024mint}, $\tau$-Bench~\citep{yao2024tau}, UltraTool~\citep{huang2024planning}, and ToolRet~\citep{shi2025retrieval}), which focus on tool selection, parameter alignment, complex long-tail invocation, and retrieval robustness; and task-oriented benchmarks (e.g., AgentBench~\citep{liu2024agentbench}, GAIA~\citep{mialon2024gaia}, WebArena~\citep{zhou2024webarena}, and TaskBench~\citep{shen2024taskbench}), which emphasize end-to-end planning, task decomposition, and completion rates in open environments. Although recent efforts like HammerBench~\citep{wang2025hammerbench} have extended evaluation to function calls in real-world mobile scenarios, mainstream benchmarks remain constrained by Outcome Bias, relying predominantly on final success rates or trajectory-level metrics. They critically lack fine-grained process annotations for multi-turn interactions, such as cross-turn context inheritance, coreference resolution, dynamic intent evolution, task atomization, and semantic tool mapping. To address this, we introduce MTDTool, a fine-grained benchmark for multi-turn mobile interactions. By providing multi-level annotations covering dialogue intent evolution chains and ranked tool candidates, it establishes a new paradigm for fine-grained process evaluation in complex mobile contexts, offering a standardized basis for quantifying the impact of task decomposition quality on tool retrieval efficacy.

\section{Conclusion}

In this work, we reveal that existing reinforcement learning-based task decomposition methods are prone to reward hacking due to ill-designed reward signals, leading to repetitive sub-task decomposition and severe OOD generalization failure. To address these issues, we propose PCTD, a framework that synergizes counterfactual rewards to cut off spurious correlations and preference rewards to enforce structural quality, fundamentally resolving the aforementioned challenges. Furthermore, we release MTDTool, a fine-grained benchmark for evaluating task decomposition in mobile multi-turn interactions. Extensive experiments show that PCTD effectively eliminates redundant decomposition, significantly outperforming SOTA methods in retrieval accuracy, decomposition quality, and OOD robustness.


\bibliography{iclr2026_conference}
\bibliographystyle{iclr2026_conference}

\clearpage
\appendix
\startcontents[appendix]

\section*{Appendix}
\addcontentsline{toc}{section}{Appendix}

\begin{center}
    \large\bfseries Contents
\end{center}

\printcontents[appendix]{l}{1}{
    \setcounter{tocdepth}{2}
}

\clearpage
\section{Baselines}
\label{baseline}

\subsection{Closed-Source LLMs}
\begin{itemize}[left=0pt]
    \item \textbf{Qwen3 Series}~\citep{yang2025qwen3} is a series of LLMs developed by the Alibaba team. It includes the general-purpose language model Qwen3, the advanced reasoning-specific model Qwen3-Reasoner, and the agent-oriented model Qwen3-Agent. The Qwen3 series of models have been pretrained on significantly larger and higher-quality datasets with extended context windows. In addition, it achieves state-of-the-art performance in complex tasks such as autonomous agent planning, multi-step mathematical reasoning, and precise instruction following.
    \item \textbf{ChatGPT-5.0} is a large-scale language model developed by OpenAI based on the Transformer architecture. It demonstrates strong capabilities in natural language understanding, reasoning, and text generation, enabling it to handle complex dialogue and knowledge-intensive tasks. 
    \item \textbf{DeepSeek-V4 Series} is a state-of-the-art open-weight large language model built on a Mixture-of-Experts (MoE) architecture with auxiliary-loss-free load balancing and FP8 mixed-precision training. It achieves superior reasoning and instruction-following performance while maintaining high computational efficiency through sparse activation and optimized inference kernels.
    \item \textbf{MiniMax-M2.5} is a high-performance proprietary large language model featuring an optimized Transformer architecture with enhanced long-context processing and multi-modal alignment capabilities. 
   
\end{itemize}

\subsection{Prompting Methods}
\begin{itemize}[left=0pt]
    \item \textbf{Q2E}~\citep{wang2023query2doc} is a semantic query expansion mechanism based on Large Language Models (LLMs). For short and ambiguous user queries, Q2E leverages the reasoning capabilities of LLMs to reconstruct them into content-rich pseudo-documents, thereby improving the relevance ranking performance of retrieval systems without requiring additional supervision signals.
    \item \textbf{ReInvoke}~\citep{chen2024re} proposes a zero-shot tool retrieval paradigm based on Large LLMs, the essence of which is to decompose user intent into tool-executable semantic units. By reformulating raw queries into structured instructions aligned with tool descriptions, Re-invoke achieves fine-grained mapping from abstract requirements to specific tool parameters. This approach effectively bridges the semantic gap between user queries and tool interfaces without requiring fine-tuning, thereby enhancing tool retrieval performance in complex multi-task scenarios.
    \item \textbf{ToolReAGt}~\citep{brauns2025toolreagt} constructs an iterative retrieval framework based on the ReAct paradigm, core to which is the mandatory embedding of tool retrieval within the Thought-Action-Observation loop. For each sub-task, the Agent generates a tailored query via Thought and mandatorily invokes the retriever to obtain candidate tools. Subsequently, in the Observation phase, it evaluates matching relevance; if no suitable tool is found, it leverages feedback to enter the next iteration for query refinement. 
    \item \textbf{PLUTO}~\citep{huang2024pluto} employs a post-retrieval planning and editing mechanism. It initially retrieves a candidate set of tools based on the user query, and subsequently refines the results through a two-step process leveraging LLMs. First, a planning module filters the most relevant tool subset from the candidates and determines their invocation order according to the current task context. Second, an editing module instantly corrects and adapts the documentation or parameter examples of the selected tools to eliminate semantic discrepancies between the original descriptions and specific task requirements. 
\end{itemize}

\subsection{RL-Based Method}
\begin{itemize}[left=0pt]
    \item \textbf{ToolQP}~\citep{fang2026beyond} reformulates tool retrieval as an iterative planning process and incorporates Reinforcement Learning with Verifiable Rewards (RLVR) to optimize the planning policy. During inference, the model decomposes abstract intents into structured sub-goals with specific parameters and preconditions via sub-goal decomposition, generating targeted queries accordingly. In the training phase, TOOLQP establishes a closed-loop optimization mechanism: it utilizes downstream tool retrieval metrics (e.g., Recall@K or MRR) as verifiable reward signals to directly guide the parameter updates of the planner. 
\end{itemize}

\subsection{Re-ranking methods}
\begin{itemize}[left=0pt]
\item \textbf{Qwen3-Embedding series}~\citep{zhang2025qwen3}is a high-performance text embedding model built upon the Qwen3 dense Transformer architecture, optimized specifically for semantic representation and retrieval tasks. It employs contrastive learning with hard negative mining and multi-stage training to achieve superior performance on MTEB benchmarks while maintaining efficient inference through knowledge distillation from larger teacher models.
\item \textbf{gte-Qwen2-1.5B-instruct}~\citep{li2023towards} is a compact text embedding model based on the Qwen2-1.5B dense Transformer architecture, fine-tuned specifically for semantic retrieval and representation tasks. It achieves competitive MTEB performance through contrastive learning with hard negative mining and instruction-aware training, while maintaining high inference efficiency suitable for resource-constrained deployment scenarios.
\item  \textbf{e5-mistral-7b-instruct}~\citep{wang2024improving} is a text embedding model built upon the Mistral-7B dense Transformer architecture, fine-tuned with instruction-aware contrastive learning for semantic retrieval tasks. It achieves strong performance on MTEB benchmarks by leveraging the base model's reasoning capabilities and optimizing representation quality through hard negative mining and multi-stage training.
\end{itemize}

\begin{table}[htbp]
\centering
\caption{Evaluation Dimensions and Penalty Rules of the Process Reward Model (PRM).}
\label{tab:prm_dimensions}
\small
\renewcommand{\arraystretch}{1.3}
\begin{tabular}{p{2.8cm} p{4.5cm} p{4.5cm}}
\toprule
\textbf{Dimension} & \textbf{Description} & \textbf{Penalty Rule} \\
\midrule
Completeness & Whether all valid intents, core entities, and constraints from the original utterance are included without omission. & Severe penalty for missing intents, entities, or constraints. \\
Accuracy & Whether the user's true intent is precisely restored without semantic deviation. & Severe penalty for semantic misinterpretation or non-executable tasks. \\
Coreference Resolution & Pronouns (e.g., ``it'', ``that'') must be replaced with explicit entities to ensure standalone readability. & Moderate penalty for unresolved coreferences. \\
Standardized Expression & Colloquialisms and filler words are removed; output should be concise and standardized instructions. & Minor penalty for non-standard or redundant expressions. \\
Contextual Consistency & Historical information is correctly inherited across turns; modifications and conflicts are properly handled. & Severe penalty for incorrect context handling. \\
\bottomrule
\end{tabular}
\end{table}

\section{Implementation Details}
\label{ap:implementation_details}
Our PCTD framework is trained in two stages. We apply LoRA adaptation (rank=64, alpha=128) to Qwen3-4B and Qwen3-8B base models initialized from SFT checkpoints. The training configuration uses a batch size of 128, a PPO mini-batch size of 64, and a PPO micro-batch size of 8 per GPU. The maximum prompt length is set to 1024 and the maximum response length to 128.

During GRPO rollout, we sample 8 candidate responses per prompt ($n=8$) and employ vLLM as the inference engine with a GPU memory utilization of 0.6. The multi-signal reward consists of two components: a preference reward with weight $w_{pref}$ and a counterfactual  reward with weight $1 - w_{pref}$. The preference reward is provided by a Qwen2.5-1.5B reward model, We set \(w_{\mathrm{pref}}=0.5\) to balance the preference reward and the
counterfactual retrieval-gain reward. Within \(R_{\mathrm{cf}}\), we set
\(\lambda_{\mathrm{rank}}=\lambda_{\mathrm{cov}}=1.0\) unless otherwise specified.

During both training and evaluation, we use a Qwen3-Embedding-4B based tool recall service to retrieve the top-10 tools for each query. Following standard tool-recall practice, each decomposed subtask independently retrieves its top-$K$ candidate tools, and the results are merged via round-robin ranking with deduplication.

\section{Evaluation of Task Decomposition Quality}
\label{ap: GSB_evalution}
\subsection{GSB Evaluation Process and Scheme}
To comprehensively evaluate the model's performance in multi-turn intent rewriting and task decomposition, we established a General Scoring Benchmark (GSB) based on the \textit{LLM-as-Judge} paradigm. 
Adopting an \textbf{absolute scoring} mode, we randomly sampled 300 instances each from In-Domain and Out-of-Domain datasets. 
Three state-of-the-art closed-source LLMs (Qwen3.6-Plus, GLM-5.1, and DeepSeek-v4-flash) served as independent judges, with their assessments aggregated via majority or weighted voting to mitigate individual model biases. 
The evaluation framework comprises two core dimensions, \textit{Rewriting Quality} and \textit{Decomposition Quality}, each assigned a 50\% weight. 
Specifically, Rewriting Quality is fine-grainedly assessed based on conciseness-fidelity (0.2), context coherence (0.3), and multi-task handling capability (0.5); 
while Decomposition Quality focuses on granularity appropriateness (0.5), logical consistency (0.2), and completeness (0.3). 
Finally, we quantified the performance gap between our method (Ours) and mainstream closed-source baselines by aggregating the weighted 5-point scale scores across all dimensions.

\begin{table}[t]
\centering
\small
\caption{GSB evaluation results on In-Domain and Out-of-Domain datasets. Scores are on a 5-point scale. \textbf{Bold} indicates the best performance.}
\label{tab:gsg_results}
\begin{tabular}{lcc}
\toprule
\textbf{Model} & \textbf{In-Domain} & \textbf{Out-of-Domain} \\
\midrule
\multicolumn{3}{l}{\textit{Closed-Source LLMs}} \\
MiniMax-M2.5 & 4.19 & 4.17 \\
DeepSeek-v4-pro & 4.27 & 4.25 \\
ChatGPT-5.0 & \textbf{4.36} & 4.34 \\
\midrule
\multicolumn{3}{l}{\textit{Ours (RLVR)}} \\
Ours w/ Preference Reward & 4.01 & 4.00 \\
\textbf{Ours method-RL} & \textbf{4.36} & \textbf{4.37} \\
\bottomrule
\end{tabular}
\end{table}

\subsection{Analysis of GSB Evaluation Results}
As shown in Table~\ref{tab:gsg_results}, the GSB evaluation demonstrates that PCTD achieves state-of-the-art performance in multi-turn intent rewriting and task decomposition. Our method significantly outperforms mainstream closed-source models, including MiniMax-M2.5 and DeepSeek-v4-pro, across both In-Domain and Out-of-Domain evaluations. Notably, in the more challenging OOD scenario, PCTD surpasses the strongest baseline ChatGPT-5.0 with a score of 4.37 versus 4.34, while achieving a comparable score of 4.36 in the In-Domain setting, validating its superior generalization capability and robustness. Furthermore, ablation studies confirm the critical role of the counterfactual reasoning mechanism: compared to the variant using only the preference reward (4.01/4.00), the complete PCTD method yields score improvements of 0.35 and 0.37 on In-Domain and OOD datasets, respectively. This substantial gain indicates that quantifying marginal causal gains via counterfactual rewards effectively severs spurious correlation shortcuts, compelling the model to learn genuine decomposition logic and fundamentally resolving the OOD generalization failure prevalent in traditional RL-based agents.

\section {Further Analysis of Dataset}

\subsection{Statistics}
\label{ap: detail_dataset_info}
The constructed dataset covers seven vertical tool categories, including daily life, travel, utilities, device control (system or application control), reminders, communication, and audio-video entertainment. Based on domain relevance, the dataset is divided into In-Domain and Out-of-Domain subsets: the In-Domain subset includes three verticals (daily life, travel, and utilities) involving 36 tools, while the Out-of-Domain subset covers the remaining four verticals with 201 tools. The dataset comprises a total of 2,917 training samples, and the test set consists of 6,058 In-Domain samples and 5,201 Out-of-Domain samples. During the retrieval evaluation phase, the full tool pool is employed, and only the \texttt{tool\_description} field is used for text embedding representation.

\subsection{Dataset distribution information}
\begin{figure}[t]
    \centering
    \includegraphics[width=\linewidth]{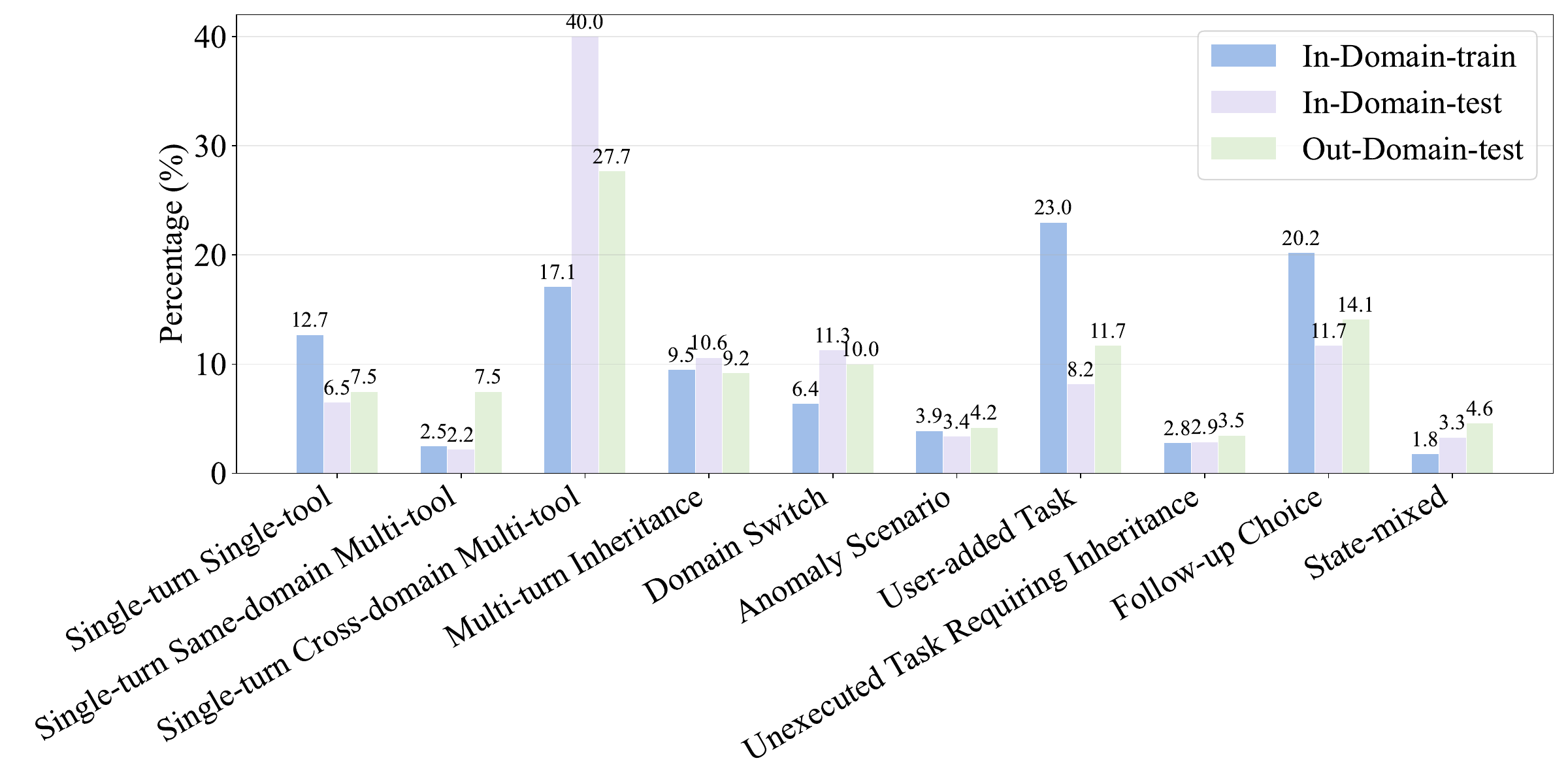}
    \caption{Dataset distribution (grouped percentage).}
    \label{ap:distribution_grouped_pct}
\end{figure}

\begin{figure}[t]
    \centering
    \includegraphics[width=\linewidth]{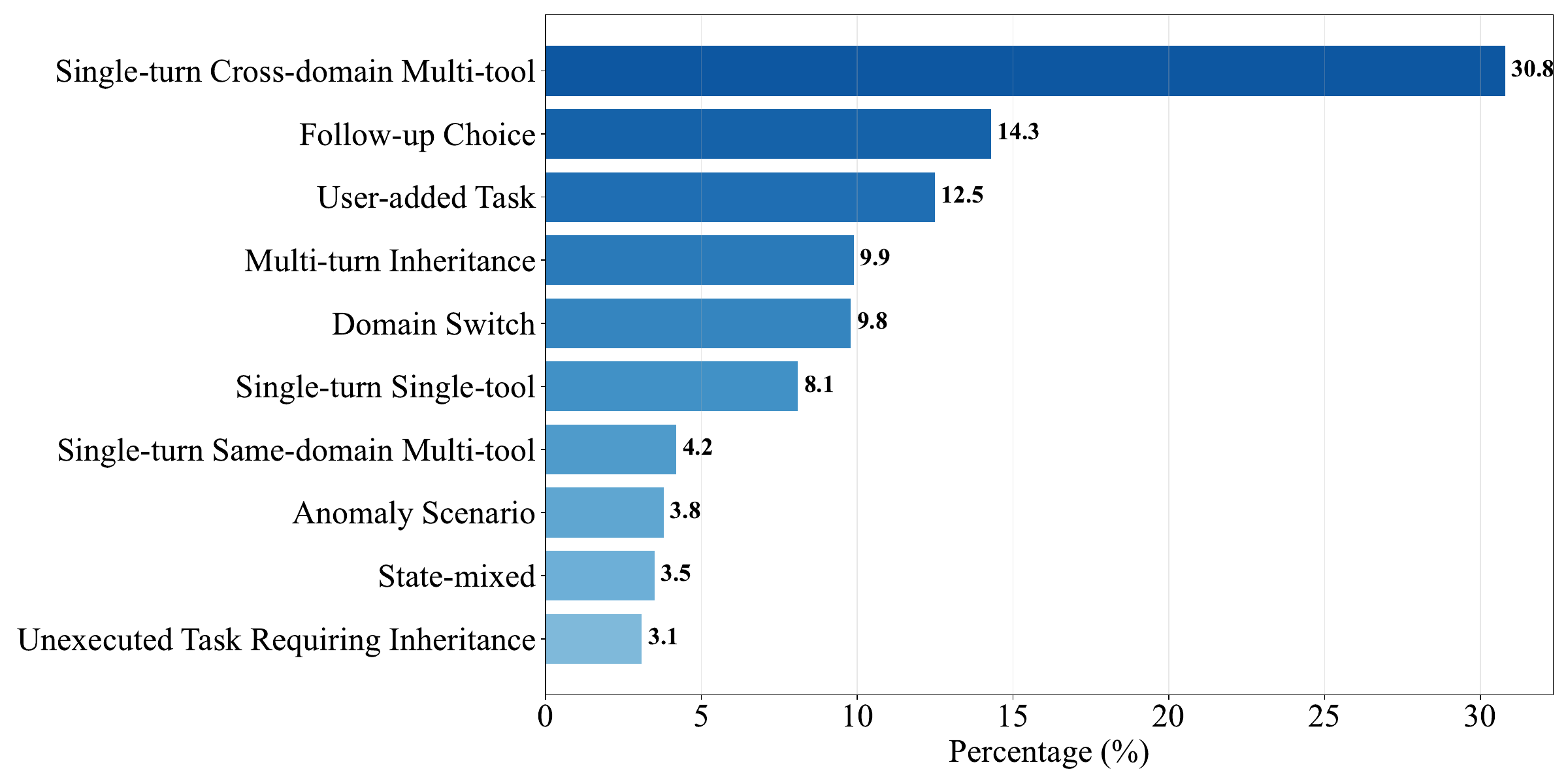}
    \caption{Overall dataset distribution.}
    \label{ap:distribution_total_hbar}
\end{figure}

We categorize the dataset into two major groups based on the dialogue interaction structure and task execution states: Single-turn and Multi-turn scenarios. On this basis, we further define several specific scenario types to systematically cover diverse interaction patterns that may arise during tool usage. The detailed classification results are presented in Table~\ref{ap:dialogue}. 
As illustrated in Figure~\ref{ap:distribution_grouped_pct}, the proportions of different categories in the training set and test set are presented in detail. Figure~\ref{ap:distribution_total_hbar} further shows the distribution of different categories across the entire dataset.

In single-turn scenarios, the user request is completed within a single dialogue turn. This category includes three types: Single-turn Single Tool, Single-turn Multi-tool (Same Category), and Single-turn Cross-category Multi-tool. Specifically, Single-turn Single Tool refers to tasks that can be completed by invoking only one tool in a single interaction, representing the most basic form of tool usage. Single-turn Multi-tool (Same Category) describes situations where multiple tools are required within a single turn, but these tools belong to the same functional category, such as multiple tools related to travel services. Single-turn Cross-category Multi-tool refers to cases where multiple tools from different domains or categories are required within one turn, for example, simultaneously checking the weather and booking a hotel.

In multi-turn scenarios, the dialogue consists of multiple interaction turns, and the execution state of tasks may evolve or be inherited across turns. We further define several common multi-turn interaction types, including Domain Switching, Exception Scenario, Unexecuted Task Inheritance, Follow-up Selection, User-added Task, Mixed State Type, and Multi-turn Inheritance. Domain Switching refers to situations where the current turn abruptly shifts to a completely new topic or domain that is unrelated to the previous turn. Exception Scenario describes cases where abnormal execution states occur during task execution, such as failures, cancellations, re-executions, or rollbacks. Unexecuted Task Inheritance indicates that a silent task from the previous turn is still running in the background and its state must be tracked or inherited in the current turn. Follow-up Selection refers to interactions where the assistant asks the user for missing information or clarification, or where the user selects among multiple options. User-added Task describes situations in which the user introduces a new task during the dialogue that is unrelated to the previous task. Mixed State Type refers to dialogues that simultaneously involve three or more of the above multi-turn interaction types (excluding multi-turn inheritance). Finally, Multi-turn Inheritance represents the normal continuation of an unfinished task from the previous turn into the current turn.
Through this categorization, the dataset aims to comprehensively cover interaction patterns ranging from basic tool invocation to complex multi-turn task management, thereby providing a more fine-grained foundation for evaluating and analyzing model capabilities.

\begin{table}[ht]
\centering
\caption{Comparison of MTDTool with existing tool-use benchmarks across different capability dimensions.}
\setlength{\tabcolsep}{1pt}
\begin{tabular}{lcccccc}
\hline
Dimension & Ours & Hammer & ToolRet & APIBench & ToolBench & API-Bank \\
\hline
Single-turn Single Tool & $\checkmark$ & $\checkmark$ & $\checkmark$ & $\checkmark$ & $\checkmark$ & $\checkmark$ \\
Single-turn Multiple Tools & $\checkmark$ & $\times$ & $\checkmark$ & $\times$ & $\checkmark$ & $\checkmark$ \\
Single-turn Cross-domain Multiple Tools & $\checkmark$ & $\times$ & $\checkmark$ & $\times$ & $\checkmark$ & $\checkmark$ \\
Multi-turn Inheritance & $\checkmark$ & $\checkmark$ & $\times$ & $\times$ & $\checkmark$ & $\checkmark$ \\
Switching & $\checkmark$ & $\checkmark$ & $\times$ & $\times$ & $\times$ & $\times$ \\
Exception Scenarios & $\checkmark$ & $\times$ & $\times$ & $\times$ & $\times$ & $\times$ \\
User Follow-up Tasks & $\checkmark$ & $\times$ & $\times$ & $\times$ & $\times$ & $\times$ \\
Unfinished Task Inheritance & $\checkmark$ & $\times$ & $\times$ & $\times$ & $\times$ & $\times$ \\
Follow-up Selection & $\checkmark$ & $\checkmark$ & $\times$ & $\times$ & $\times$ & $\times$ \\
Mixed State Types & $\checkmark$ & $\times$ & $\times$ & $\times$ & $\times$ & $\times$ \\
\hline
\end{tabular}
\end{table}

\begin{algorithm}[H]
\caption{Counterfactual and Preference Rewards for Task Decomposition}
\label{alg:cf_gdpo}

\begin{algorithmic}[1]
\REQUIRE SFT policy $\pi_{\theta}$, old policy $\pi_{\theta_{\mathrm{old}}}$, reference policy $\pi_{\mathrm{ref}}$, preference model $s_{\phi}$, tool retrieval service $\mathcal{S}_{\mathrm{tr}}$, weights $w_{\mathrm{pref}}, \lambda_{\mathrm{rank}}, \lambda_{\mathrm{cov}}$
\ENSURE Optimized policy $\pi_{\theta}$

\FOR{each training step}
    \STATE Sample $n$ decompositions $\{y_t^{(j)}\}_{j=1}^{n} \sim \pi_{\theta_{\mathrm{old}}}(\cdot \mid x_t)$ for each prompt $x_t=(H_t,q_t)$

    \STATE Counterfactual reward: for each $y_t^{(j)}$,
    \STATE \quad Retrieve tools for decomposition and raw context: $\mathcal{L}_{\mathrm{atom}}^{K,(j)} \leftarrow \mathcal{S}_{\mathrm{tr}}(y_t^{(j)}, K)$, $\mathcal{L}_{\mathrm{raw}}^K \leftarrow \mathcal{S}_{\mathrm{tr}}(x_t, K)$
    \STATE \quad $\Delta_{\mathrm{rank}}^{(j)} \leftarrow \mathrm{NDCG@}K(\mathcal{L}_{\mathrm{atom}}^{K,(j)}, \mathcal{G}_t) - \mathrm{NDCG@}K(\mathcal{L}_{\mathrm{raw}}^K, \mathcal{G}_t)$
    \STATE \quad $\Delta_{\mathrm{cov}}^{(j)} \leftarrow \mathbb{I}[\mathcal{G}_t \subseteq \mathcal{L}_{\mathrm{atom}}^{K,(j)}] - \mathbb{I}[\mathcal{G}_t \subseteq \mathcal{L}_{\mathrm{raw}}^K]$
    \STATE \quad $R_{\mathrm{cf}}^{(j)} \leftarrow \lambda_{\mathrm{rank}}\max(0,\Delta_{\mathrm{rank}}^{(j)}) + \lambda_{\mathrm{cov}}\Delta_{\mathrm{cov}}^{(j)}$

    \STATE Preference reward: $R_{\mathrm{pref}}^{(j)} \leftarrow \sigma\!\left(s_{\phi}(x_t,y_t^{(j)}) - s_{\phi}(x_t,y_t^\star)\right)$

    \STATE Multi-signal fusion: $R^{(j)} \leftarrow w_{\mathrm{pref}} \cdot R_{\mathrm{pref}}^{(j)} + (1-w_{\mathrm{pref}}) \cdot R_{\mathrm{cf}}^{(j)}$

    \STATE GRPO advantage: normalize within the $n$-candidate group,
    \STATE \quad $\hat{A}^{(j)} \leftarrow \dfrac{R^{(j)} - \frac{1}{n}\sum_{k=1}^{n} R^{(k)}}{\operatorname{std}(\{R^{(k)}\}_{k=1}^{n})+\epsilon_{\mathrm{std}}}$

    \STATE Policy update: minimize
    \STATE \quad $L \leftarrow -\mathbb{E}\!\Big[\min\!\big(\rho^{(j)} \hat{A}^{(j)},\, \operatorname{clip}(\rho^{(j)}, 1{-}\epsilon_{\mathrm{clip}}, 1{+}\epsilon_{\mathrm{clip}}) \hat{A}^{(j)}\big)\Big] + \beta_{\mathrm{KL}} \cdot \mathrm{KL}\big(\pi_{\theta} \,\|\, \pi_{\mathrm{ref}}\big)$
    \STATE \quad where $\rho^{(j)} = \pi_{\theta}(y_t^{(j)} \mid x_t) / \pi_{\theta_{\mathrm{old}}}(y_t^{(j)} \mid x_t)$
\ENDFOR

\RETURN $\pi_{\theta}$
\end{algorithmic}
\end{algorithm}

\begin{table}[ht]
\centering
\caption{Definitions of dialogue scenario labels used in the dataset.}
\label{ap:dialogue}
\setlength{\tabcolsep}{4pt}
\small
\begin{tabular}{@{}llp{6.8cm}@{}}
\toprule
Scenario & Label & Brief Description \\
\midrule
\multirow{3}{*}{\centering Single-turn} 
& Single-turn Single Tool 
& A single-turn dialogue that completes a simple task using only one tool. \\[4pt]

& Single-turn Multi-tool (Same Category) 
& A single-turn dialogue that uses multiple tools within the same category (e.g., multiple ``travel'' tools). \\[4pt]

& Single-turn Cross-category Multi-tool 
& A single-turn dialogue that uses multiple tools across different categories (e.g., checking weather and booking a hotel). \\
\midrule

\multirow{7}{*}{\centering Multi-turn} 
& Domain Switching 
& In a multi-turn dialogue, the current turn switches to a completely new topic or domain unrelated to the previous turn. \\[4pt]

& Exception Scenario 
& A multi-turn dialogue containing abnormal execution states such as failure, cancellation, re-execution, or rollback. \\[4pt]

& Unexecuted Task Inheritance 
& A silent task from the previous turn is still running in the background and its state must be inherited in the current turn. \\[4pt]

& Follow-up Selection 
& The assistant asks follow-up questions for missing details, or the user makes a selection during multi-turn dialogue. \\[4pt]

& User-added Task 
& The user introduces a new task unrelated to the task from the previous turn during multi-turn dialogue. \\[4pt]

& Mixed State Type 
& A multi-turn dialogue simultaneously including three or more of the above scenarios (excluding multi-turn inheritance). \\[4pt]

& Multi-turn Inheritance 
& The current turn normally continues executing an unfinished task from the previous turn in multi-turn dialogue. \\
\bottomrule
\end{tabular}
\end{table}

\begin{table}[h]
\centering
\small
\setlength{\tabcolsep}{4pt}
\renewcommand{\arraystretch}{1.2}
\caption{Hyperparameter analysis of reward weights $w_{pref}$. Performance metrics (N@10, C@10) are reported in percentage (\%). Repetition rates are reported as raw values.}
\label{tab:hyperparameter}
\begin{tabular}{c c | c c c | c c c}
\hline
\multirow{2}{*}{$w_{pref}$} & \multirow{2}{*}{$1-w_{pref}$} & \multicolumn{3}{c|}{In Domain} & \multicolumn{3}{c}{Out Domain} \\
\cmidrule(lr){3-5} \cmidrule(lr){6-8}
 & & N@10 & C@10 & Duplication Rate (\%) & N@10 & C@10 & Duplication Rate (\%) \\
\hline
0 & 1 & 90.65 & 86.18 & 5.24 & 78.59 & 75.79 & 1.60 \\
\rowcolor{blue!8} \textbf{0.2} & \textbf{0.8} & \textbf{91.19} & \textbf{87.08} & 0.71 & \textbf{82.74} & \textbf{79.77} & 1.35
 \\
0.5 & 0.5 & 90.50 & 86.63 & 0.70 & 81.85 & 79.50 & 1.00 \\
0.8 & 0.2 & 86.55 & 80.59 & 0.36 & 77.74 & 73.24 & 0.10 \\
1 & 0 & 89.49 & 85.70 & 0.41 & 78.93 & 75.77 & 0.17 \\
\hline
\end{tabular}
\end{table}

\section{Further Analysis}

\subsection{When might PCTD fail?}
\label{ap: when_fails}
Although PCTD improves performance through the joint reward mechanism, its effectiveness depends on the balance between the counterfactual reward $R_{\mathrm{cf}}$ and the preference reward $R_{\mathrm{pref}}$. Table~\ref{tab:hyperparameter} reports the effect of the preference weight $w_{\mathrm{pref}}$ on retrieval performance and repetition rate, revealing the potential failure modes of PCTD under extreme reward configurations.
When $w_{\mathrm{pref}}=0$, the model is optimized solely by the counterfactual reward. Although it achieves a high In-Domain N@10 of 90.65, its Out-of-Domain performance drops to 78.59, and the repetition rate increases to 5.24. This result suggests that counterfactual retrieval feedback alone is insufficient to guarantee high-quality task decomposition. Without structural preference constraints, the policy may still exploit retrieval metrics through redundant or repetitive decomposition patterns, leading to weaker OOD generalization.

As $w_{\mathrm{pref}}$ increases, the preference reward gradually provides stronger structural regularization. The repetition rate decreases substantially from 5.24 to 0.17, while the Out-of-Domain N@10 reaches its best value of 82.74 when $w_{\mathrm{pref}}=0.2$. However, when $w_{\mathrm{pref}}$ becomes too large, e.g., $w_{\mathrm{pref}}=1$, the model relies only on preference supervision. Although this setting effectively suppresses repetition, the In-Domain N@10 decreases to 89.49, indicating that overly strong structural regularization may weaken the model's ability to optimize retrieval-specific objectives.

These results show that the main failure risk of PCTD comes from an imbalance between the two reward signals. When $w_{\mathrm{pref}}\rightarrow 0$, the lack of preference supervision makes the model prone to repetitive decomposition and OOD degradation. When $w_{\mathrm{pref}}\rightarrow 1$, excessive preference supervision may over-regularize the policy and sacrifice retrieval accuracy. Therefore, PCTD performs best when the preference reward and the counterfactual reward are properly balanced, enabling the model to learn decompositions that are both structurally well-formed and effective for tool retrieval.

\begin{table}[h]
\centering
\small
\setlength{\tabcolsep}{4pt}
\renewcommand{\arraystretch}{1.2}
\caption{Retrieval performance on the MTDTool benchmark using different retrievers.}
\label{ap: retrievers}
\begin{tabular}{lcccc}
\hline
 & \multicolumn{2}{c}{In-Domain} & \multicolumn{2}{c}{Out-of-Domain} \\
\cline{2-5}
Method & N@10 & C@10 & N@10 & C@10 \\
\hline

\textit{\textbf{Qwen3-Embedding-4B}} \\
Base Retriever & 68.35 & 61.79 & 50.57 & 42.22 \\
Ours & 91.19 & 87.08 & 82.74 & 79.77 \\
$\Delta$ (\%) & 22.84 & 25.29 & 32.17 & 37.55 \\
\hline

\textit{\textbf{gte-Qwen2-1.5B-instruct}} \\
Base Retriever & 61.93 & 54.74 & 48.01 & 38.20 \\
Ours & 87.01 & 83.93 & 81.40 & 80.97 \\
$\Delta$ (\%) & 25.08 & 29.19 & 33.39 & 42.77 \\
\hline

\textit{\textbf{bge-large-zh-v1.5}} \\
Base Retriever & 52.06 & 39.58 & 40.98 & 29.69 \\
Ours & 81.67 & 76.01 & 79.89 & 78.10 \\
$\Delta$ (\%) & 29.61 & 36.43 & 38.91 & 48.41 \\
\hline

\textit{\textbf{Qwen3-Embedding-0.6B}} \\
Base Retriever & 64.42 & 58.26 & 48.43 & 39.84 \\
Ours & 88.54 & 86.60 & 84.27 & 82.43 \\
$\Delta$ (\%) & 24.12 & 28.34 & 35.84 & 42.59 \\
\hline

\textit{\textbf{e5-mistral-7b-instruct}} \\
Base Retriever & 55.73 & 48.09 & 40.58 & 34.07 \\
Ours & 84.10 & 80.44 & 71.45 & 65.87 \\
$\Delta$ (\%) & 28.37 & 32.35 & 30.87 & 31.80 \\
\hline

\textit{\textbf{Fine-tuned Qwen3-Embedding-4B}} \\
Base Retriever & 93.37 & 87.97 & 72.20 & 64.10 \\
Ours & 93.31 & 89.81 & 82.85 & 84.06 \\
$\Delta$ (\%) & -0.06 & 1.84 & 10.65 & 19.96 \\
\hline

\end{tabular}
\end{table}
\subsection{Performance of PCTD Across Different Retrievers}
\label{ap: pctd_with_retrievers}
Table~\ref{ap: retrievers} presents the zero-shot generalization performance of PCTD across different retrievers after training on MTDTool. The experiments cover a range of mainstream embedding models from lightweight (0.6B) to large-scale (7B+), as well as a Qwen3-Embedding-4B model fine-tuned on downstream tasks. Key analyses and conclusions are as follows:

1.  \textbf{Strong Retriever-Agnostic Capability:} PCTD achieves significant improvements across all tested base retrievers. Notably, for the smaller-parameter models \texttt{Qwen3-Embedding-0.6B} and \texttt{bge-large-zh-v1.5}, In-Domain N@10 increases by 24.12\% and 29.61\%, respectively, while Out-of-Domain C@10 surges by 42.59\% and 48.41\%. This demonstrates that the atomic sub-tasks generated by PCTD possess extremely high semantic signal-to-noise ratios, effectively compensating for the semantic understanding limitations of lightweight retrievers and enabling them to approach or even surpass unoptimized strong retrievers.
2.  \textbf{More Pronounced Robustness Enhancement in OOD Scenarios:} Compared to In-Domain performance, the improvement margin of PCTD in OOD scenarios is generally 10--20 percentage points higher. For instance, on \texttt{gte-Qwen2-1.5B}, OOD C@10 improves by 42.77\%, far exceeding the In-Domain gain of 29.19\%. This indicates that by cutting off spurious correlations via counterfactual rewards, the decomposition logic learned by the model exhibits stronger causal universality rather than merely memorizing tool co-occurrence patterns in the training set, thereby generating precise retrieval queries even when facing unseen tools.
3.  \textbf{Complementary Advantage over Fine-Tuned Retrievers:} Even against the downstream fine-tuned \texttt{Fine-tuned Qwen3-Embedding-4B}, PCTD remains competitive. Although the absolute gain narrows on this specific fine-tuned model due to its high adaptation to the task distribution (In-Domain N@10 drops slightly by 0.06\%, but C@10 still rises by 1.84\%; OOD N@10 increases by 10.65\%), this precisely highlights that the core value of PCTD lies in \textbf{optimizing the semantic quality at the input end}. When the retriever is already sufficiently strong, PCTD shifts from ``score boosting'' to ``tackling hard cases,'' continuing to unlock potential in OOD difficult examples; whereas for un-tuned generic retrievers, PCTD plays a decisive role in achieving substantial performance leaps.

In summary, these results compellingly demonstrate that PCTD is not merely an optimization strategy tailored for specific retrievers, but a \textbf{universal semantic enhancement paradigm}. By transforming ambiguous instructions into high signal-to-noise atomic sub-tasks, it fundamentally resolves the semantic mismatch problem in agent tool retrieval, enabling even lightweight or generic retrieval models to achieve high-precision tool localization in complex multi-turn interactions and unknown tool scenarios.

\begin{table}[t]
    \centering
    \caption{Definition of Tool Execution States and Typical Transition Logic. The transition logic follows the two-stage process $d \sim P_{\mathrm{todo}}(s) \rightarrow s' \sim P_{\mathrm{task}}(d)$.}
    \label{tab:state_definitions}
    \resizebox{\columnwidth}{!}{
    \begin{tabular}{l c l p{4.5cm} p{5.5cm}}
        \toprule
        \textbf{Category} & \textbf{ID} & \textbf{State Name} & \textbf{Core Semantics} & \textbf{Typical Transition Logic ($P_{\mathrm{todo}} \rightarrow P_{\mathrm{task}}$)} \\
        \midrule
        \multirow{3}{*}{Regular Execution} 
        & $s_1$ & Success & Task completed & Null (0.95) / Revocation (0.05) $\rightarrow$ Remove / Reverse Op \\
        & $s_2$ & In Progress & Task ongoing & Retry (0.8) / Cancel (0.2) $\rightarrow$ Success / Failure \\
        & $s_3$ & Failure & Abnormal termination & Retry (0.9) / Cancel (0.1) $\rightarrow$ Retry / Terminate \\
        \addlinespace[0.5em]
        \multirow{2}{*}{Interaction Clarification} 
        & $s_4$ & Clarification & Missing key info & Supplementation (1.0) $\rightarrow$ Re-execute after user input \\
        & $s_5$ & Selection & Multi-option confirm & Selection Confirm (1.0) $\rightarrow$ Execute corresponding branch \\
        \addlinespace[0.5em]
        \multirow{6}{*}{Intent Management} 
        & $s_6$ & Silence & Demand not triggered & Maintain silence / Transition to new task \\
        & $s_7$ & Cancellation & Explicit refusal & Null (1.0) $\rightarrow$ Tool exits context \\
        & $s_8$ & Revocation & Rollback operation & Null (1.0) $\rightarrow$ Execute reverse tool \\
        & $s_9$ & Retry & Retry request & Success (0.95) / Failure (0.05) \\
        & $s_{10}$ & Supplementation & Active completion & Success (1.0) $\rightarrow$ Continue subsequent tasks \\
        & $s_{11}$ & Null & Exit dialogue & -- \\
        \bottomrule
    \end{tabular}%
    }
\end{table}

\begin{figure*}[t]
\centering
\begin{minipage}{0.95\linewidth}
\hrule
\vspace{0.5em}
\noindent\textbf{System:} You are a real human phone user interacting with an AI voice assistant. Generate a natural spoken command.
\vspace{0.5em}

\noindent\textbf{User:}
\begin{quote}
\noindent Below is the list of tools available to the assistant (for your reference only; you must trigger them implicitly):
\begin{verbatim}
available_tools:
[
{"tool_name": "create_reminder",
"description": "Create, query, delete and modify reminders",
"actions": ["create", "query", "delete"],
"objects": ["time", "title", "repeat rule"]},
{"tool_name": "weather_service",
"description": "Query weather for a specified city and date",
"actions": ["query weather"],
"objects": ["city", "date"]},
{"tool_name": "calendar_service",
"description": "Query, create, modify and delete calendar events",
"actions": ["query", "create", "delete"],
"objects": ["title", "start time", "end time", "location"]}
]
\end{verbatim}

\noindent Conceive a natural human voice command, strictly following these rules:

\noindent (1) \textbf{Order}: express demands in the exact order of the tool list.

\noindent (2) \textbf{One-tool-one-request}: exactly 3 tools $\Rightarrow$ exactly 3 demands.

\noindent (3) \textbf{Colloquial conversion}: do not copy official terminology verbatim.

\noindent (4) \textbf{No long sentences}: do not cram all demands into one sentence.

\vspace{0.3em}
\noindent\textbf{State guidance} (injected when a tool is assigned the \textit{Ask-Detail} state):
\begin{quote}
Tool \texttt{weather\_service} is in state \textbf{Ask-Detail}. You must \textbf{deliberately omit exactly one} critical information slot (e.g., city) in the corresponding demand. The omission must be achieved through natural colloquial phrasing, with no explanatory language.
\end{quote}

\noindent\textbf{Linguistic style} (one feature randomly sampled per turn):
\begin{quote}
Use \textit{fuzzy quantifiers and relative time}---avoid exact numbers or absolute timestamps; use expressions like "a bit", "later", "in a couple of days" in exactly one demand.
\end{quote}
\end{quote}

\vspace{0.5em}
\noindent\textbf{Expected output:} Set a reminder for me to pick up the kid this afternoon. Oh, by the way, what's the weather like tomorrow? And while you're at it, check if I've got any meetings on Wednesday.
\vspace{0.3em}
\hrule
\end{minipage}
\caption*{First-turn user prompt for instruction generation. Variable blocks (tool list, state guidance, linguistic style) are dynamically injected based on the state-machine configuration.}
\label{ap:user_prompt}
\end{figure*}

\begin{figure*}[t]
\centering
\begin{minipage}{0.95\linewidth}
\hrule
\vspace{0.5em}
\noindent\textbf{System:} You are an AI voice assistant. Generate a natural, helpful response based on the user's command and backend execution states.
\vspace{0.5em}

\noindent\textbf{User:} ``Set a reminder for me to pick up the kid this afternoon. Oh, by the way, what's the weather like tomorrow? And while you're at it, check if I've got any meetings on Wednesday.''
\vspace{0.5em}

\noindent\textbf{Assistant:}
\begin{quote}
\noindent\textit{Backend execution states:}
\begin{lstlisting}[breaklines=true,basicstyle=\ttfamily\small]
- Tool: Create, query, delete and modify reminders
  State: [Success]
- Tool: Query weather for a specified city and date
  State: [Ask-Detail]
  Follow-up direction: city
- Tool: Query, create, modify and delete calendar events
  State: [Silent]
\end{lstlisting}

\noindent (2) \textbf{Coherent delivery}: blend multiple task feedback into one fluent response; no bullet-point reporting.

\noindent (3) \textbf{No tool leakage}: never reveal tool names, codes, or backend terminology.
\end{quote}

\vspace{0.3em}
\noindent\textbf{Expected output:} ``Your reminder to pick up the kid this afternoon has been set. By the way, which city would you like me to check the weather for tomorrow?''
\vspace{0.3em}
\hrule
\end{minipage}
\caption*{First-turn assistant prompt. Responses are strictly driven by execution states; the \textit{Silent} tool is completely absent from the output.}
\label{ap:assistant_prompt_example}
\end{figure*}

\begin{figure*}[t]
\centering
\begin{minipage}{0.95\linewidth}
\hrule
\vspace{0.5em}
\noindent\textbf{System:} You are a real human phone user. Generate the next-turn spoken command, building on the prior conversation.
\vspace{0.5em}

\noindent\textit{Dialogue history:}
\begin{quote}
\textbf{User:} Set a reminder to pick up the kid this afternoon. Oh, by the way, what's the weather like tomorrow? And check if I've got any meetings on Wednesday.''\\ \textbf{Assistant:} Your afternoon reminder is set. Which city would you like me to check the weather for?''
\end{quote}

\noindent\textbf{User (next turn):}
\begin{quote}
\noindent\textit{This turn requires:}
\begin{itemize}[nosep,leftmargin=*]
\item Answer the assistant's follow-up question about the missing city.
\item Introduce new demands corresponding to the tools below.
\end{itemize}

\noindent\textit{Tools and target states:}
\begin{lstlisting}[breaklines=true,basicstyle=\ttfamily\small]
["weather_service", "calendar_service", "wifi_service"]
States: [Ask-Detail, Select, Silent]
\end{lstlisting}

\noindent\textbf{Rules:}
\begin{enumerate}[nosep,leftmargin=*]
\item \textbf{Supplement}: provide the previously omitted city name naturally; do not repeat the full original instruction.
\item \textbf{Selection}: choose one of the assistant's offered options; output only the option itself.
\item \textbf{Silent}: completely skip this tool---no mention, hint, or summary.
\item \textbf{Order}: demands must follow the tool list order.
\item \textbf{One-tool-one-request}: exactly 3 tools $\Rightarrow$ 3 demands.
\item \textbf{No confirmation words}: no "okay", "thanks", etc.
\end{enumerate}

\noindent\textit{Linguistic style (one demand only):} Use \textit{inverted sentence}---front-load the object before the action, e.g., ``That QQ Music, open it for me.''
\end{quote}

\vspace{0.3em}
\noindent\textbf{Expected output:} ``Beijing. The first meeting on Wednesday, cancel that. And turn on the Wi-Fi.''
\vspace{0.3em}
\hrule
\end{minipage}
\caption*{Next-turn user prompt (context-inheritance mode). The user must simultaneously answer the assistant's prior follow-up, make a selection, introduce new demands, and silently skip one tool---all in a single utterance.}
\end{figure*}

\begin{figure*}[t]
\centering
\begin{minipage}{0.95\linewidth}
\hrule
\vspace{0.5em}
\noindent\textbf{System:} You are an AI voice assistant. Generate a natural response based on the user's next-turn command and backend execution states.
\vspace{0.5em}

\noindent\textbf{User:} ``Beijing. The first meeting on Wednesday, cancel that. And turn on the Wi-Fi.''
\vspace{0.5em}

\noindent\textbf{Assistant:}
\begin{quote}
\noindent\textit{Backend execution states:}
\begin{lstlisting}[breaklines=true,basicstyle=\ttfamily\small]

Tool: Query weather for a specified city and date State: [Success]
Tool: Query, create, modify and delete calendar events State: [Ask-Detail] Follow-up direction: specific date to cancel
Tool: Toggle Wi-Fi connectivity State: [Executing] \end{lstlisting}
\noindent\textbf{Response rules:}
\begin{enumerate}[nosep,leftmargin=*]
\item \textbf{State-driven}:
\begin{itemize}[nosep]
\item \textit{Success}: confirm completion.
\item \textit{Ask-Detail}: ask a specific question for the missing information.
\item \textit{Executing}: inform the user the task is being processed.
\item \textit{Silent}: completely skip, as if the demand never existed.
\end{itemize}
\item \textbf{Coherent}: blend all feedback into one fluent response.
\item \textbf{No tool leakage}: never reveal tool names or backend terms.
\end{enumerate}
\end{quote}

\vspace{0.3em}
\noindent\textbf{Expected output:} ``I've checked the weather for Beijing---it looks clear tomorrow. Which specific meeting on Wednesday would you like me to cancel? Meanwhile, I'm turning on the Wi-Fi for you.''
\vspace{0.3em}
\hrule
\end{minipage}
\caption*{Next-turn assistant prompt. Identical rule structure to the first-turn version; only the injected user query and tool states differ.}
\end{figure*}

\begin{figure*}[t]
\centering
\begin{minipage}{0.95\linewidth}
\hrule
\vspace{0.5em}
\noindent\textbf{System:} You are a multi-turn rewriting and task decomposition expert. Decompose user commands into single-turn, single-task units.
\vspace{0.5em}

\noindent\textit{Dialogue history:}
\begin{quote}
\textbf{User:} Set a reminder to pick up the kid this afternoon. Oh, by the way, what's the weather like tomorrow? And check if I've got any meetings on Wednesday.''\\ \textbf{Assistant:} Your afternoon reminder is set. Which city would you like me to check the weather for?''
\end{quote}

\noindent\textbf{Current query:} ``Beijing. The first meeting on Wednesday, cancel that. And turn on the Wi-Fi.''
\vspace{0.5em}

\noindent\textbf{Task definition:}
\begin{enumerate}[nosep,leftmargin=*]
\item \textbf{Rewriting}: resolve pronouns and omissions using context. E.g., (Q: open Bluetooth)(A: done)(Q: turn it off)'' $\rightarrow$ turn off Bluetooth''.
\item \textbf{Multi-task splitting}: separate compound commands into atomic tasks. E.g., (Q: open Wi-Fi, Bluetooth and dark mode)'' $\rightarrow$ open Wi-Fi$<br>$open Bluetooth$<br>$enable dark mode''.
\end{enumerate}

\noindent\textbf{Output format:} separate tasks with \texttt{<br>}.

\vspace{0.5em}
\noindent\textbf{Expected output:}
\begin{quote}
``Query weather for Beijing$<br>$cancel the first Wednesday meeting$<br>$turn on Wi-Fi''
\end{quote}
\vspace{0.3em}
\hrule
\end{minipage}
\caption*{Task decomposition prompt. Invoked independently at each turn to produce an atomic task sequence $\mathcal{A}_t$, forming the middle tier of the three-level annotation structure (context $\rightarrow$ atomic tasks $\rightarrow$ candidate tools).}
\end{figure*}

\begin{figure*}[t]
\centering
\begin{minipage}{0.98\textwidth}
\hrule
\vspace{0.4em}

\noindent\small\textbf{Round 1 (Initial)} --- Tools: 
\texttt{create\_reminder}, \texttt{weather\_service}, \texttt{calendar\_service} \\
States: [Success, Ask-Detail, Silent]

\vspace{0.3em}
\noindent\textbf{User:} ``Set a reminder to pick up the kid this afternoon. 
Oh, by the way, what's the weather like tomorrow? 
And check if I've got any meetings on Wednesday.''

\vspace{0.2em}
\noindent\textbf{Task split:} \\
``Set reminder to pick up kid this afternoon'' \\
``Query tomorrow's weather'' \\
``Check Wednesday meetings''

\vspace{0.2em}
\noindent\textbf{Assistant:} 
``Your afternoon reminder is set. 
Which city would you like me to check the weather for tomorrow?'' 
\hfill{\footnotesize\textit{(note: calendar\_service $\rightarrow$ Silent, completely absent)}}

\vspace{0.6em}

\noindent\small\textbf{Round 2 (Inherit)} --- Tools: 
\texttt{weather\_service}, \texttt{calendar\_service}, \texttt{wifi\_service} \\
States: [Ask-Detail, Select, Executing]

\vspace{0.3em}
\noindent\textbf{User:} 
``Beijing. The first meeting on Wednesday, cancel that. 
And turn on the Wi-Fi.''

\vspace{0.2em}
\noindent\textbf{Task split:} \\
``Query weather for Beijing'' \\
``Cancel first Wednesday meeting'' \\
``Turn on Wi-Fi''

\vspace{0.2em}
\noindent\textbf{Assistant:} 
``Got it---tomorrow in Beijing will be partly cloudy, around 18 degrees. 
Your Wednesday meeting has been cancelled. 
Meanwhile, I'm enabling Wi-Fi for you.''

\vspace{0.6em}

\noindent\small\textbf{Round 3 (Inherit)} --- Tools: 
\texttt{navigation\_service}, \texttt{calendar\_service}, \texttt{create\_reminder} \\
States: [Success, Retry, Ask-Detail]

\vspace{0.3em}
\noindent\textbf{User:} 
``Navigate to the company. That meeting thing, try again. 
Also remind me to bring my laptop---oh wait, make it two reminders, 
one for the laptop and one for the charger.''

\vspace{0.2em}
\noindent\textbf{Task split:} \\
``Navigate to company'' \\
``Retry creating Wednesday meeting'' \\
``Remind to bring laptop'' \\
``Remind to bring charger''

\vspace{0.2em}
\noindent\textbf{Assistant:} 
``Route to the company is ready, about 25 minutes. 
I'm retrying the meeting creation now. 
What time should I set the laptop reminder for?''

\vspace{0.6em}

\noindent\small\textbf{Round 4 (Switch)} --- Tools: 
\texttt{music\_service}, \texttt{bluetooth\_service} \\
States: [Success, Success]

\vspace{0.3em}
\noindent\textbf{User:} 
``Play some light music. Oh, and connect the headphones.''

\vspace{0.2em}
\noindent\textbf{Task split:} \\
``Play light music'' \\
``Connect Bluetooth headphones''

\vspace{0.2em}
\noindent\textbf{Assistant:} 
``Playing a light music playlist. 
Your Bluetooth headphones are now connected.''

\vspace{0.4em}
\hrule
\end{minipage}

\caption*{End-to-end multi-turn dialogue construction example. Round~1 initializes states; Rounds~2--3 inherit context (answering follow-ups, retrying failed tasks, introducing new tools); Round~4 switches to a new topic. Each round independently invokes user prompt, task decomposition, and assistant prompt.}
\label{ap:full_prompt}
\end{figure*}

\

\end{document}